\documentclass[useAMS,usenatbib]{mn2e}
\usepackage{aas_macros,graphicx,times,multirow,amsmath,amssymb,color,longtable}
\title[Periodic signals from the Circinus region]{Periodic signals from the Circinus region: two new cataclysmic variables and the ultraluminous X-ray source candidate GC\,X--1}
\author[P.~Esposito et al.] {P.~Esposito,$^{1,2}$\thanks{E-mail: paoloesp@iasf-milano.inaf.it} G.~L.~Israel,$^{3}$ D. Milisavljevic,$^{2}$ M.~Mapelli,$^{4,5}$  L.~Zampieri,$^4$ L.~Sidoli,$^1$ \newauthor G.~Fabbiano$^{2}$ and G. A. Rodr\'{\i}guez Castillo$^{3}$ 
\smallskip\\
$^1$INAF--Istituto di Astrofisica Spaziale e Fisica Cosmica - Milano, via E. Bassini 15, I-20133 Milano, Italy\\
$^2$Harvard--Smithsonian Center for Astrophysics, 60 Garden Street, Cambridge, MA 02138, USA\\
$^3$INAF--Osservatorio Astronomico di Roma, via Frascati 33, I-00040 Monteporzio Catone, Italy\\
$^4$INAF--Osservatorio Astronomico di Padova, vicolo dell'Osservatorio 5, I-35122 Padova, Italy\\
$^5$INFN, Sezione di Milano Bicocca, piazza della Scienza 3, I-20126, Milano, Italy
}
\date{Accepted 2015 June 18.  Received 2015 June 9; in original form 2015 May 14} \pagerange{\pageref{firstpage}--\pageref{lastpage}} \pubyear{2015}

\def\LaTeX{L\kern-.36em\raise.3ex\hbox{a}\kern-.15em
    T\kern-.1667em\lower.7ex\hbox{E}\kern-.125emX}
\def\xmm {\emph{XMM--Newton}}
\def\cxo {\emph{Chandra}}
\def\swift {\emph{Swift}}

\def\sax {\emph{BeppoSAX}}

\def\rst {\emph{ROSAT}}

\def\hst {\emph{HST}}
\def\cgx {\mbox{CG\,X--1}}
\def\ip {\mbox{CXO\,J141430}}
\def\cv {\mbox{CXOU\,J141332}}
\def\flux {\mbox{erg cm$^{-2}$ s$^{-1}$}}
\def\lum {\mbox{erg s$^{-1}$}}
\def\nh {$N_{\rm H}$}

\begin{document}

\label{firstpage}
\maketitle
\begin{abstract}
The examination of two 2010 \cxo\ ACIS exposures of the Circinus galaxy resulted in the discovery of two pulsators: CXO\,J141430.1--651621 and CXOU\,J141332.9--651756. We also detected 26-ks pulsations in \cgx, consistently with previous measures. For $\sim$40 other sources, we obtained limits on periodic modulations. In CXO\,J141430.1--651621, which is $\sim$2~arcmin outside the Circinus galaxy, we detected  signals at $6120\pm1$~s and $64.2\pm0.5$~ks. In the longest observation, the source showed a flux of $\approx$$1.1\times10^{-13}$~\flux\ (absorbed, 0.5--10~keV) and the spectrum could be described by a power-law with photon index $\Gamma\simeq1.4$. From archival observations, we found that the luminosity is variable by $\approx$50~per~cent on time-scales of weeks--years. The two periodicities pin down CXO\,J141430.1--651621 as a  cataclysmic variable of the intermediate polar subtype. The period of CXOU\,J141332.9--651756 is $6378\pm3$~s. It is located inside the Circinus galaxy, but the low absorption indicates a Galactic foreground object. The flux was $\approx$$5\times10^{-14}$~\flux\ in the \cxo\ observations and showed $\approx$50~per~cent variations on weekly/yearly scales; the spectrum is well fit by a power law with $\Gamma\simeq0.9$. These characteristics and the large modulation suggest that CXOU\,J141332.9--651756 is a magnetic cataclysmic variable, probably a polar. For \cgx, we show that if the source is in the Circinus galaxy, its properties are consistent with a Wolf--Rayet plus black hole binary. We consider the implications of this for ultraluminous X-ray sources and the prospects of Advanced LIGO and Virgo. In  particular, from the current sample of WR--BH systems we estimate an upper limit to the detection rate of stellar BH--BH mergers of $\sim$16~yr$^{-1}$.
\end{abstract}

\section{Introduction}

The \cxo\ ACIS Timing Survey at Brera And Rome astronomical observatories project (CATS\,@\,BAR; Israel et al., in preparation) is a Fourier-transform-based systematic search for new pulsating sources in the \cxo\ Advanced CCD Imaging Spectrometer (ACIS; \citealt{garmire03}) public archive. As of 2015 April 30, 10,282 ACIS Timed Exposure observations have been examined and $\sim$457,000 sources were detected. Data taken with gratings or in Continuous-Clocking mode were not considered. The $\sim$93,600 light curves of sources with more than 150 photons were searched for coherent signals with an algorithm based on that of \citet{israel96}. The limit of 150 counts is related to the intrinsic ability of the  Fourier transform to detect a signal with 100~per~cent modulation at a minimum confidence level of 3.5$\sigma$ in $10^{5}$--$10^{6}$ trials. CATS\,@\,BAR has so far discovered 43 new certain X-ray pulsators; see \citet{eis13,eisrc13,eism13} for the first results and \citet{eis14,esposito15} for our analogous \swift\ project.

In this paper, we report on the CATS\,@\,BAR results for the galaxy ESO 97--G13 (the `Circinus galaxy', hereafter CG; \citealt{freeman77}) and its surroundings in the Circinus constellation. CG is a nearby Seyfert~II active galaxy that lies close to the plane of our own Galaxy (J2000 Galactic coordinates: $l=311\fdg3$, $b= -03\fdg8$; distance $d=4.2$~Mpc; \citealt{tully09}). The CG contains hydrogen-rich star forming regions in the inner spiral arms and, due to its closeness, offers a good opportunity to study its population of X-ray sources \citep{bauer01,sambruna01}, which includes several ultraluminous X-ray sources (ULXs, \citealt{winter06,mapelli10,walton13}; see \citealt{mushotzky04,fabbiano06,zampieri09,feng11} for reviews on ULXs).

The CG was observed several times with \cxo, but it was in two long Timed Exposure observations carried out in late 2010 that CATS\,@\,BAR pinpointed two new X-ray pulsators in the Circinus region: \ip\ and the uncatalogued \cv. The pipeline detected also \cgx\ (CXOU\,J141312.3--652013), whose emission is modulated at $\sim$7~h. \cgx\ is not a new pulsator, but the long-standing debate about its nature \citep{sw01,bauer01,weisskopf04} prompted us to include it in our study.

The plan of the paper is as follows. In Section\,\ref{observations} we give details on the X-ray observations used in our study. The rest of the paper is divided into two main parts. The first one focuses on the new pulsators and comprises Sections \ref{timing_analysis} to \ref{discussion}. In Section\,\ref{timing_analysis} we describe the timing analysis that led to the discovery of the new pulsators and allowed us also to set upper limits on the pulsations for dozens of other X-ray sources. The detailed study of \ip\ is presented in Section \ref{polar}, that of \cv\ in Section \ref{cataclysmic}. To study these sources, we used also data from \xmm\ and optical observations taken with the VLT Survey Telescope (VST). The optical observations and their analysis are described in Section\,\ref{vst_data}. The nature of \ip\ and \cv\ is discussed in Section\,\ref{discussion}.

The second part of the paper is dedicated to \cgx. In Section\,\ref{cgx1} we recall the main facts about this source. The results from the analysis of the 2010 \cxo\ data, which were not used before to study \cgx, are summarised in Section\,\ref{gcx_analysis}. In Section\,\ref{cgx_discussion}, we propose that \cgx\  might be a Wolf--Rayet plus black hole (WR--BH) binary system, and consider the implications of this possibility for ULXs and for the prospects of detection of gravitational radiation from BH--BH mergers. A summary with conclusive remarks follows in Section\,\ref{summary}.

\section{X-ray observations}\label{observations}

\begin{figure*}
\centering
\resizebox{\hsize}{!}{\includegraphics[angle=0]{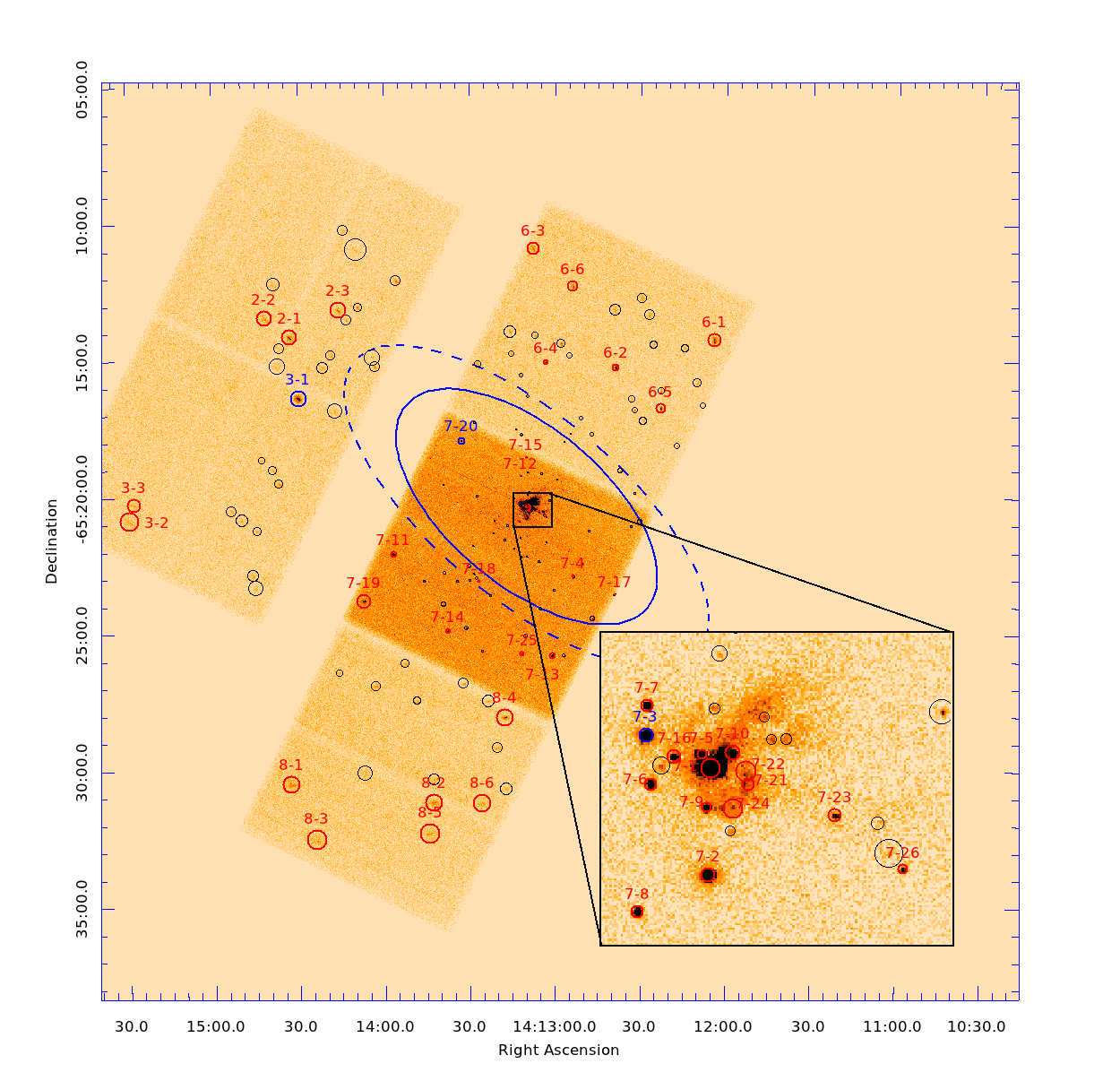}}
\caption{\label{circinus_ds9} \cxo/ACIS image of the Circinus field. North is top and east is left; each CCD subtends an $8\farcm4\times8\farcm4$ square on the sky and the zoomed area is $1\farcm2\times1'$. Circles mark the detected sources. Red circles indicate sources with $>$150 counts for which we performed a timing analysis, blue circles the sources from which periodic signals were detected. The labelled sources (with the ACIS CCD number and an identification number) are those for which either signals were detected (source 7--3 is \cgx, 7--20 is \cv, and 3--1 is \ip) or an upper limit on the pulsed fraction could be placed (see Fig.\,\ref{ulimits}). The solid and dashed ellipses indicate the size of the CG from 90~per~cent total $B$ light and total infrared (2MASS) magnitude, respectively (from the NASA/IPAC Extragalactic Database, NED, see \mbox{http://ned.ipac.caltech.edu/}). Source 7--1 is the CG's active galactic nucleus (AGN), 7--2 is the young supernova remnant candidate in the CG (CG\,X--2; \citealt{bauer01}); other notable sources are the ultraluminous X-ray sources 7--5 = Circinus ULX3, 7--16= Circinus ULX4, 7--4 = Circinus XMM1 = ULX5, 7--17 = Circinus XMM2 (while Circinus XMM3 is undetected; we used the nomenclature of \citealt{winter06,gladstone13,walton13}).}
\end{figure*}

All the observations used in this work are summarised in Table\,\ref{obslog}. The most important observations are the one in which CATS\,@\,BAR detected the new pulsators and its companion, 12823 and 12824, marked with a hash mark in Table\,\ref{obslog}. They were carried out in a week in 2010 December to study the central region of the CG \citep{mingo12}. Their combined exposure is $\sim$190~ks. In both cases, three ACIS-S and  two ACIS-I CCDs were used in full frame mode, ensuring a wide coverage over the Circinus region.
\begin{table*}
\begin{minipage}{16.cm}
\centering \caption{Summary of the X-ray observations used in this work. The hash marks and the star indicate the \cxo\ and \xmm\ observations from which most of the information was obtained.} \label{obslog}
\begin{tabular}{@{}lccccc}
\hline
Satellite & Instrument & Obs.\,ID  & Date & Exp. & Mode$^{a}$\\
&  & & & (ks) & \\
\hline
\cxo & ACIS-23678 & 355 & 2000 Jan 16 & 1.3 & TE FAINT (3.24\,s)\\
\cxo & ACIS-235678 & 356 & 2000 Mar 14 & 24.7 & TE FAINT (3.24\,s)\\
\cxo & ACIS-235678& 2454  & 2001 May 02 & 4.4 & TE FAINT (3.24\,s)\\
\emph{XMM}~* & pn / MOS\,1 / MOS\,2 & 0111240101 & 2001~Aug~6--7& 100.6 / 104.1 / 104.0& FF (73.4\,ms) / FF (2.6\,s) / LW (0.9/2.7\,s) \\
\cxo & ACIS-456789 & 10873 & 2009 Mar 01 & 18.1 & TE HETG VFAINT (2.04\,s)\\
\cxo & ACIS-456789 & 10850 & 2009 Mar 03 & 13.8 & TE HETG VFAINT (2.04\,s)\\
\cxo & ACIS-456789 & 10872 & 2009 Mar 04 & 16.5 & TE HETG VFAINT (2.04\,s)\\
\cxo~\# & ACIS-23678 & 12823 & 2010~Dec~17--19 & 154.4 & TE VFAINT (3.14\,s)\\
\cxo~\# & ACIS-23678 & 12824 & 2010~Dec~24 & 39.4 & TE VFAINT (3.14\,s)\\
\emph{XMM} & pn / MOS\,1 / MOS\,2 & 0656580601  & 2014 Mar 01 & 32.7 / 37.8 / 37.9 & FF (73.4\,ms) / FF (2.6\,s) / LW (2.7\,s)\\
\hline
\end{tabular}
\begin{list}{}{}
\item[$^{a}$] TE: Timed Exposure, HETG: High Energy Transmission Grating, VFAINT: Very Faint telemetry format, FF: Full Frame, LW: Large Window; the readout time is given in parentheses, for the central and peripheral CCDs in the case of the MOS\,2 in LW. 
\end{list}
\end{minipage}
\end{table*}
The \cxo\ data were processed and analysed with the \cxo\ Interactive Analysis of Observations software package (\textsc{ciao}, version 4.7; \citealt{fruscione06}) and the calibration files in \textsc{caldb} version 4.6.7. The Circinus field as imaged with \cxo\ in observation 12823 is shown in Fig.\,\ref{circinus_ds9}. In the data sets 12823 and 12824, \cgx\ and \cv\ were positioned in the back-illuminated CCD 7 (S3). The photons from these sources were accumulated within a circle with radius 1.5~arcsec and an ellipse with semi-axes 3.5 and 3~arcsec, respectively. \ip\ fell on the front-illuminated CCD 3 (I3) and the source counts were extracted from an ellipse with semi-axes of 15 and 14~arcsec. The choice of regions of different size is due to the point-spread function at the off-axis angles of the sources. For each source, the background was estimated locally, using source-free regions as close as possible to the target. The Solar system barycentre correction to the photon arrival times was applied with \textsc{axbary}. The spectra, the redistribution matrices,  and the ancillary response files were created using \textsc{specextract}.

The second most useful observation for our study is that performed with \xmm\ in 2001 August with a duration of $\sim$100~ks (obs. ID 0111240101; \citealt*{molendi03}; it is marked by a star in Table\,\ref{obslog}). We used the data collected with the European Photon Imaging Camera (EPIC), which consists of two MOS \citep{turner01} and one pn \citep{struder01} CCD detectors. 
The raw data were reprocessed using the \xmm\ Science Analysis Software (\textsc{sas}, version 14.0) and the calibration files in the \textsc{ccf} release of 2015 March. The observation suffered of intense soft-proton flares. The intervals of flaring background were located by intensity filters (see e.g. \citealt{deluca04}) and excluded from the analysis. This reduced the net exposure time by $\sim$30~per~cent in the pn back-illuminated CCDs and $\sim$10~per~cent in the MOS front-illuminated CCDs. The source photons were extracted from circles with radius of 25~arcsec for \ip\ and 15~arcsec for \cv\ (these radii were essentially imposed by CCD gaps and/or the presence of neighbouring sources) and the backgrounds from regions in the same chip as the sources. \cv\ was in the unread part of the central CCD of the MOS\,2 operated in a partial window mode, so only pn and MOS\,1 data exist for it. Photon arrival times were converted to the Solar system barycentre using the \textsc{sas} task \textsc{barycen}. The ancillary response files and the spectral redistribution matrices for the spectral analysis were generated with \textsc{arfgen} and \textsc{rmfgen}, respectively. Due to the low number of photons, we combined for each source the spectra from the available EPIC cameras and averaged the response files using \mbox{\textsc{epicspeccombine}}.

We made use of other \cxo\ and \xmm\ observations, which were reduced and analysed in a similar way; they provided only detections and flux estimates, or upper limits for the two new pulsators. Six data sets were collected with \cxo\ from 2000 to 2009 with various instrumental setups and durations from $\sim$1 to 25~ks, and one with \xmm\ in 2014 with exposure of $\sim$30~ks. Apart from these observations, listed in Table\,\ref{obslog}, no other \cxo\ pointing of the CG covered the positions of \ip\ or \cv, while in a $\sim$60~ks observations performed with \xmm\ in 2013 (obs. ID 0701981001), both sources fell either in gaps or at the edge of CCDs, or outside the field of view of the instruments. 

\section{CATS\,@\,BAR Timing analysis}\label{timing_analysis}

The \textsc{ciao wavedetect} routine detected 156 sources in the ACIS field of view of observation 12823; they are marked by circles in Fig.\,\ref{circinus_ds9}. The 44 sources with more than 150 photons (those marked by magenta and red circles in Fig.\,\ref{circinus_ds9}) were searched for periodic signals. 
The CATS\,@\,BAR search algorithm is based on a fast Fourier transform and takes into account also the possible presence of additional non-Poissonian noise components in the Leahy-normalised  \citep{leahy83} power spectra (see \citealt{israel96} for more details). Correspondingly, the CATS\,@\,BAR  signal threshold in a power spectrum, which takes into account the number of independent Fourier frequencies, is a function also of the local underlying noise.
For the Circinus data, the maximum frequency of the search ($\sim$0.16~Hz) is dictated by the sampling time of 3.14~s, while the longest period to which the observation is realistically sensitive (because of its duration) is $\approx$80~ks; 32\,768 frequencies were searched. The search resulted in the detection of five sources with significant signals in their power spectra. 

In two cases the power peaks were coincident with the frequencies of known spurious signals due to the spacecraft dithering pattern. The CATS\,@\,BAR pipeline automatically performs check for these artificial signals by means of the \textsc{dither\_region ciao} task.\footnote{See \mbox{http://cxc.harvard.edu/ciao/ahelp/dither\_region.html}.} Furthermore, every candidate signal is crosschecked with the CATS\,@\,BAR database of recurring signals of instrumental origin, and repeating or dubious signals are carefully inspected and rejected. 
These two sources are the supernova remnant candidate CG\,X--2 \citep{bauer01}, labelled  7--2 in Fig.\,\ref{circinus_ds9}, and Circinus XMM2, which is classified as an ULX \citep{winter06}, labelled 7--17.
In a third object, \cgx, the detected $\sim$27~ks-period modulation was already known (\citealt{bauer01}; object 7--3 in Fig.\,\ref{circinus_ds9}). This source is discussed in detail in Sections\,\ref{cgx1} to \ref{cgx_discussion}. The remaining two sources, 3--1 = \ip\ (Section\,\ref{polar}) and the uncatalogued 7--20 = \cv\ (Section\,\ref{cataclysmic}), are genuine new X-ray pulsators, as it was confirmed also by the other data sets. 

\begin{figure*}
\centering
\resizebox{.75\hsize}{!}{\includegraphics[angle=-90]{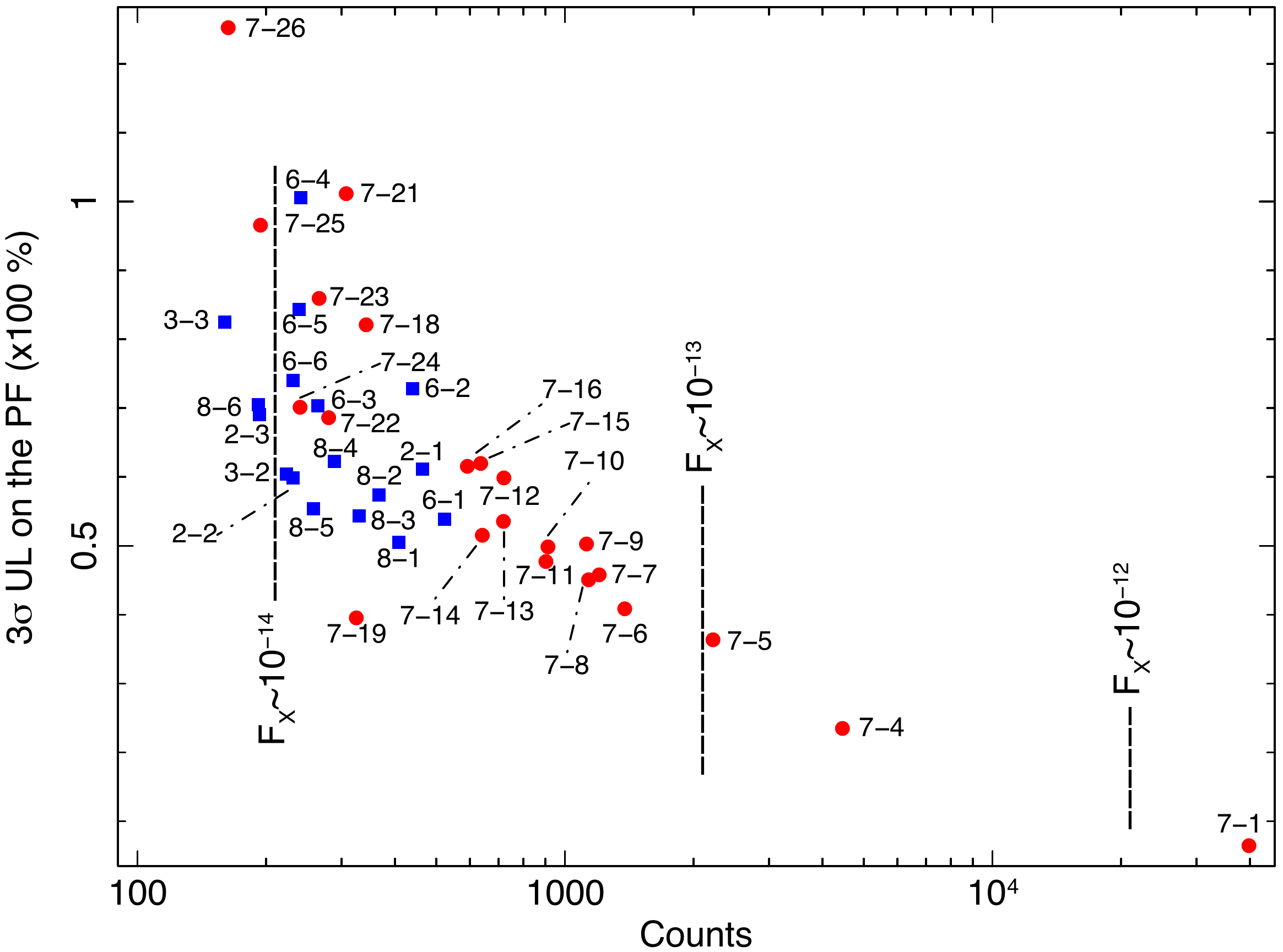}}
\caption{\label{ulimits} Upper limits on the pulsed fraction of the Circinus sources. Red circles indicate sources detected in the back-illuminated ACIS-7 (S3)
CCD, blue squares  sources from front-illuminated CCDs. The sources are identified by the same labels as in Fig\,\ref{circinus_ds9}. Dashed vertical lines  mark approximate fluxes for the S3 CCD sources. Source 7--1 is the CG's AGN; other notable sources are: 7--5 = Circinus ULX3, 7--16 = Circinus ULX4, and 7--4 = Circinus XMM1 = ULX5.}
\end{figure*}

For all the other sources with more than 150 events, a 3$\sigma$ upper limit to the pulsed fraction of any sinusoidal signal was calculated (throughout the paper, we will give 3$\sigma$ upper limits on non-detections and limits at 90~per~cent confidence level on poorly constrained quantities; all uncertainties will be given at 1$\sigma$ confidence level). The pulsed fraction was defined as the semi-amplitude of the sinusoidal modulation divided by the mean count rate. As expected, around 100--200 photons the upper limits start crossing the 100~per~cent threshold, above which no meaningful information related to any coherent signal can be inferred. For many other sources, the upper limits are not constraining. For future reference, all the results are summarised in Fig.\,\ref{ulimits}.

\section{The 1.7 / 17.8~\MakeLowercase{h} pulsator: \ip}\label{polar}

\subsection{Timing analysis}
\begin{figure*}
\centering
\resizebox{\hsize}{!}{\includegraphics[angle=-90]{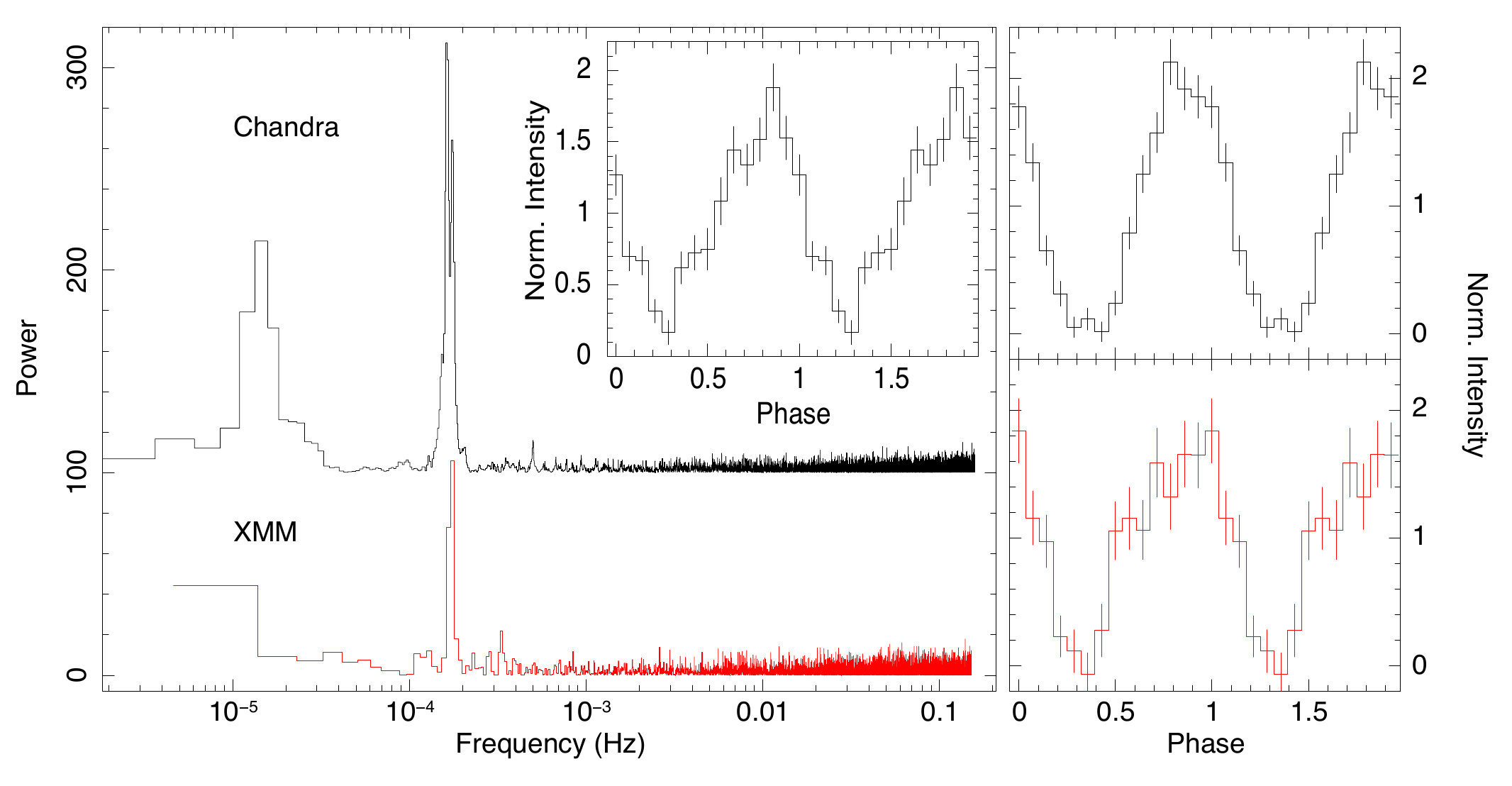}}
\caption{\label{dpsfold1414} Left panel: 0.5--10~keV power spectra of \ip\ obtained from \cxo/ACIS (top, black line, observations 12823 and 12824 combined) and \xmm/EPIC (bottom, red line) data. For displaying purposes, the \cxo\ data have been shifted in power by +100. The corresponding \cxo\ (top) and \xmm\ (bottom) light curves folded to the best periods are shown in the panels on the  right for $P_1=6.1$~ks. In the inset is shown the \cxo\ light curve folded to the longer second period $P_2=64.2$~ks.}
\end{figure*}

\ip\ is the brightest of the two new CATS\,@\,BAR pulsators. It shows two distinct periodic signals: a $\sim$100~per~cent-modulation at about 6.1~ks and another large-amplitude signal at about 64~ks. The power spectrum is shown in Fig.\,\ref{dpsfold1414}. When the 32\,768 frequencies analysed and the number of sources for which the search was carried out (44) are taken into account, both signals were detected at a confidence level larger than 10$\sigma$.

To hone the estimates of the periods, we made use of both the \cxo\ 12823 and 12824 pointings, where about 820 and 250 net counts, respectively, were collected. For the short signal, we used a phase-fitting technique and found $P_1=6\,120\pm2$~s. For the 64~ks period, the number of sampled cycles, approximately three, is too small for the phase fitting. We therefore binned the light curve to 6\,120~s, so to avoid beat-frequency signals produced by the shorter periodicity, and fit a sinusoidal function to it. The fit has a $\chi^2$ of 54 for 30 degrees of freedom (dof) and we derived the period $P_2=64.2\pm0.5$~ks. The 0.5--10~keV background-subtracted light curves folded on our best periods are shown in Fig.\,\ref{dpsfold1414}. We measured the following pulsed fractions: $100\pm4$~per~cent ($P_1=6.1$~ks; this value is to be regarded as a lower limit) and $70\pm4$~per~cent ($P_2=64.2$~ks). Within the statistical uncertainties, the shape and the pulsed fraction of both signals are energy-independent. In the soft ($<$2~keV) and hard ($>$2~keV) bands we measured pulsed fractions of $96\pm5$~per~cent and $106\pm5$~per~cent for the 6.1~ks period, and $69\pm5$~per~cent and $64\pm7$~per~cent for the 64.2~ks period.

The 6.1-ks signal is significantly detected also in the 2001 \xmm/EPIC data ($\sim$700 net counts between the three EPIC cameras), while the observation is too short for the 64.2-ks period (Fig.\,\ref{dpsfold1414}). We measured the period $P_1=6.04\pm0.04$~ks and a pulsed fraction of $88\pm12$~per~cent. \ip\ is detected with $\sim$170 net counts  in the 2014 \xmm\ pointing (pn plus MOS\,2, in the MOS\,1 the source fell in one of the failed CCDs). The short-period pulsations are clear also in that data set, but the low count statistics hampers a precise estimate of the period. Finally, \ip\ was in the field of view of \cxo\ also in the observations 355 (2000 January, 1.3-ks) and 356 (2000 March 25-ks; Table\,\ref{obslog}). In observation 355, \ip\ is detected with a dozen of photons only, and a signal-to-noise ratio $\rm SNR \ga 3$: the short duration and the very small number of photons preclude any analysis of the periodic signals. In the data set 356, the source is detected with about 90 photons ($\rm SNR> 9$). The 6.1-ks signal can be clearly observed but, similarly than in the 2014 \xmm\ observations, the uncertainty on the period is very large. 

\subsection{Spectral analysis}

\begin{figure*}
\centering
\resizebox{\hsize}{!}{\includegraphics[angle=0]{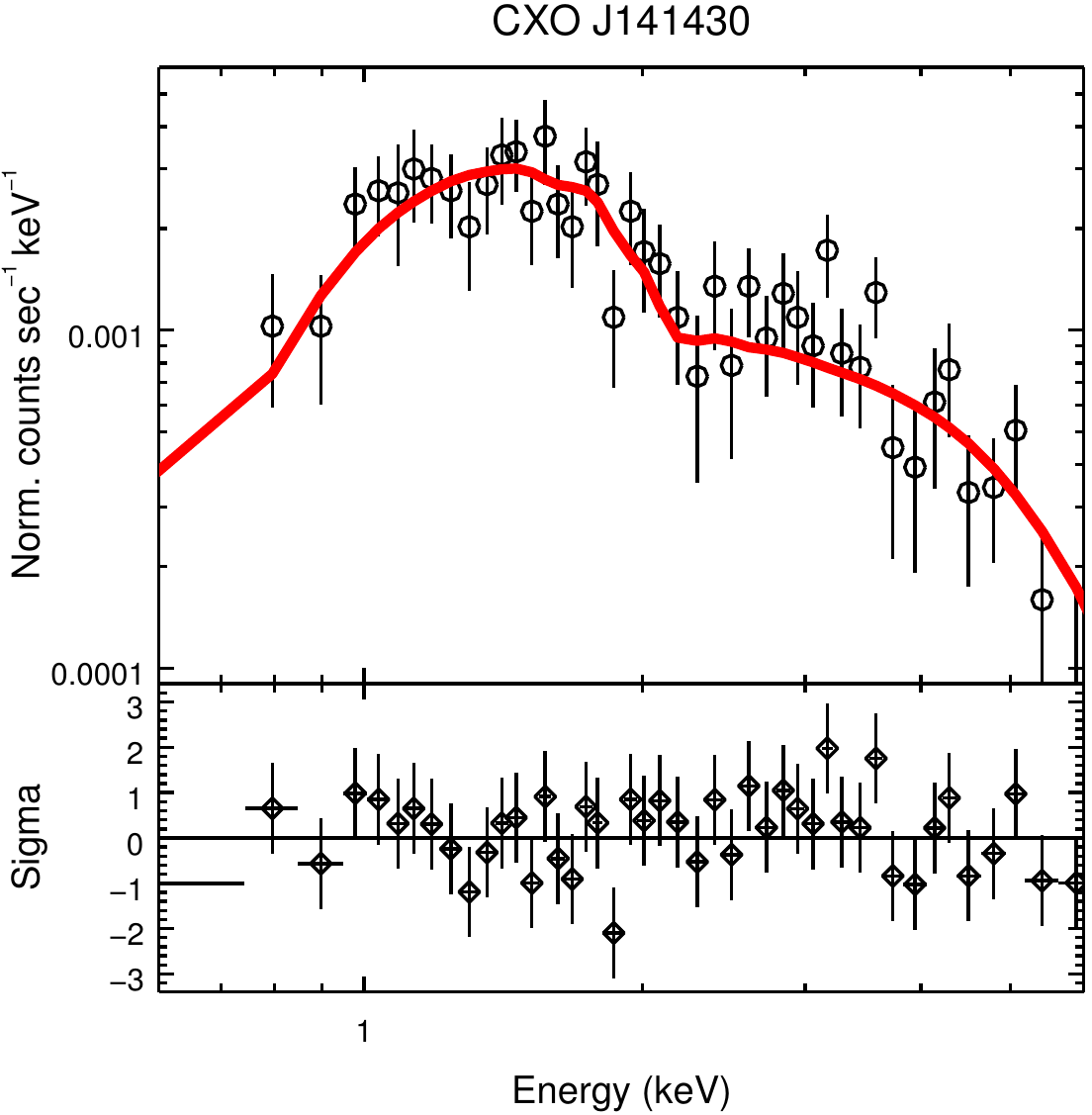}\hspace{1cm}\includegraphics[angle=0]{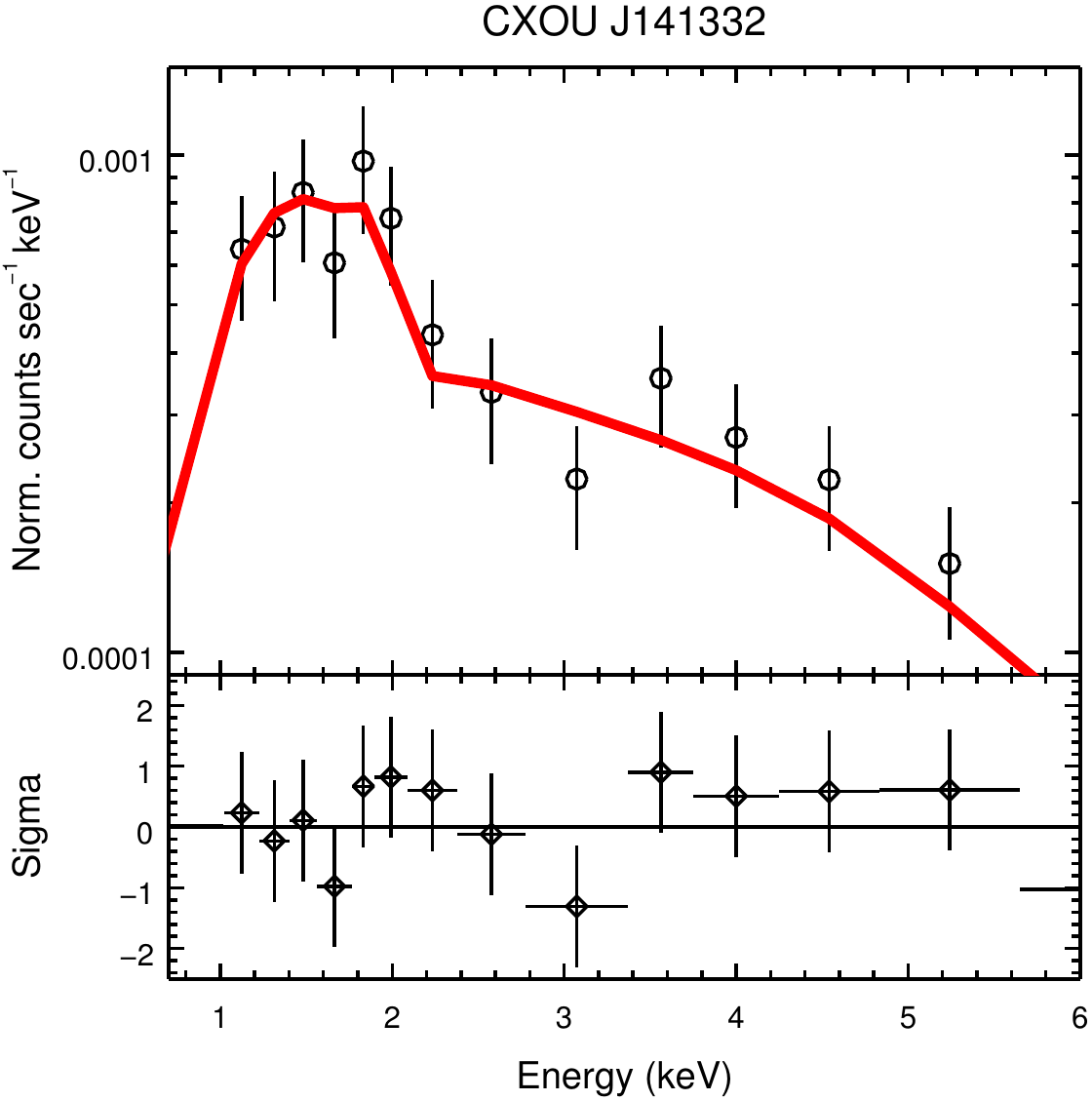}}
\caption{\label{cvs_spec} \cxo/ACIS spectra and best-fitting power-law models (red solid line) for \ip\ and \cv\ (as indicated on each panel) from observation 12823. Bottom panels: residuals in units of standard deviations.}
\end{figure*}

For the spectral analysis, we started from the long \cxo\ observation 12823. The fits were performed between 0.6 and 6~keV because of the very low signal of \ip\ outside this range. We fit to the data a power law model, a blackbody, and an optically-thin thermal bremsstrahlung. The blackbody model yielded a reduced $\chi^2_\nu=1.37$ for 40 dof with clearly structured residuals; the derived temperature is $kT=0.81\pm0.03$~keV, while for the absorption there is only an upper limit of $N_{\rm H}<0.8\times10^{22}$~cm$^{-2}$ at 90~per~cent confidence. The observed flux was $F_{\mathrm{X}}=6.8^{+0.5}_{-0.4}\times10^{-14}$~\flux\ (0.5--10~keV). The power law and the bremsstrahlung gave somewhat better fits, $\chi^2_\nu=1.27$ and $1.24$, respectively, and better residuals. The parameters of the power law fit (Fig.\,\ref{cvs_spec}, left) are $N_{\rm H}=0.36^{+0.14}_{-0.13}\times10^{22}$~cm$^{-2}$, $\Gamma=1.51^{+0.16}_{-0.15}$, and $F_{\mathrm{X}}=(1.03\pm0.09)\times10^{-13}$~\flux. For the bremsstrahlung, $N_{\rm H}=0.31^{+0.11}_{-0.10}\times10^{22}$~cm$^{-2}$, $kT=13^{+12}_{-5}$~keV, and $F_{\mathrm{X}}=0.97^{+0.09}_{-0.10}\times10^{-13}$~\flux.

In the second \cxo\ observation, the flux was $\approx$30~per~cent higher. The absorption was only poorly constrained ($<$$0.6\times10^{22}$~cm$^{-2}$ at 90~per~cent confidence for the power law and $<$$0.5\times10^{22}$~cm$^{-2}$ for the bremsstrahlung), while $kT$ and $\Gamma$ were consistent with those measured in observation 12823. We thus decided to fit the two spectra simultaneously, with the normalisations free to vary and the other parameters tied up between the data sets. The results are summarised in Table\,\ref{ip_specs}.
\begin{table*}
\begin{minipage}{16.9cm}
\centering \caption{Spectral results of \ip. Errors are at a 1$\sigma$ confidence level for a single parameter of interest.} \label{ip_specs}
\begin{tabular}{@{}lccccccc}
\hline
Model & Data & \nh$^a$ & $\Gamma$ & $kT$ & Flux$^b$ & Unabsorbed flux$^b$ & $\chi^2_\nu$ (dof) \\
 & & ($10^{22}$~cm$^{-2}$) &  & (keV) & \multicolumn{2}{c}{($10^{-13}$ \flux)} &  \\
\hline
\multirow{2}{*}{\textsc{phabs(powerlaw)} }& \cxo/12823 & \multirow{2}{*}{$0.30^{+0.13}_{-0.11}$} & \multirow{2}{*}{$1.43^{+0.14}_{-0.13}$} & \multirow{2}{*}{--} & $1.07\pm0.08$ & $1.21\pm0.07$ & \multirow{2}{*}{1.12 (52)}\\
 & \cxo/12824 &  &  & & $1.40^{+0.15}_{-0.14}$ & $1.58^{+0.13}_{-0.12}$ & \\
\hline
\multirow{2}{*}{\textsc{phabs(bremsstrahlung)}} & \cxo/12823 & \multirow{2}{*}{$0.27\pm0.10$} &  \multirow{2}{*}{--}  & \multirow{2}{*}{$18^{+24}_{-7}$} & $1.01^{+0.10}_{-0.09}$ & $1.13\pm0.08$ & \multirow{2}{*}{1.10 (52)}\\
 & \cxo/12824 & &   &  & $1.32^{+0.16}_{-0.15}$ & $1.47\pm0.14$ & \\
\hline
\textsc{phabs(powerlaw)} & \emph{XMM}/0111240101 & $<$0.3$^c$ & $1.10\pm0.15$ & -- & $0.87\pm0.08$ & $0.89^{+0.07}_{-0.06}$ & 1.02 (28) \\
\hline
\end{tabular}
\begin{list}{}{}
\item[$^{a}$] The abundances used are those of \citet*{wilms00}; \nh\ values $\approx$30~per~cent lower are derived with those by \citet{anders89}. The photoelectric absorption cross-sections are from \citet{balucinska92}.
\item[$^{b}$] In the 0.5--10~keV energy range. 
\item[$^{c}$] Upper limit at 90~per~cent confidence level. 
\end{list}
\end{minipage}
\end{table*}

The 2001 \xmm\ data flatly reject the blackbody model, with $\chi^2_\nu=2.33$ for 28 dof. The power-law provides a good fit to the data, with an observed flux  similar to that of the first \cxo\ observation. The bremsstrahlung fit was equally good. Its temperature, however, could not be constrained, as it always pegged to the highest allowed value, showing that in the \xmm\ data its curvature is indistinguishable from that of a power law. For this reason, in Table\,\ref{ip_specs} we give only the parameters derived from the power law fit. The quality of the spectrum from the 2014 \xmm\ data  is too poor for a spectral analysis. We thus used the models in Table\,\ref{ip_specs} to estimate the flux of \ip, obtaining $F_{\mathrm{X}}=(4.8\pm0.6)\times10^{-14}$~\flux\ for the power-law model and $(3.4\pm0.5)\times10^{-14}$~\flux\ for the bremsstrahlung.
Similarly, for the 2000 \cxo\ observation 356 ($\sim$90 counts in a front-illuminated ACIS-S CCDs) we evaluated a flux $F_{\mathrm{X}}=(1.1\pm0.2)\times10^{-13}$~\flux\ for the power-law model and $(8.0\pm0.9)\times10^{-14}$~\flux\ for the bremsstrahlung. In the short  pointing 355, where only a dozen of photons were detected in one of the ACIS-I CCDs, we converted the count rate into a flux with \textsc{pimms}\footnote{We used the web version of \textsc{pimms} (Portable, Interactive Multi-Mission Simulator) available at\\ \mbox{http://heasarc.gsfc.nasa.gov/cgi-bin/Tools/w3pimms/w3pimms.pl}.} and found $F_{\mathrm{X}}=(2.0\pm0.7)\times10^{-13}$~\flux\ for the power-law model and $(1.5\pm0.5)\times10^{-13}$~\flux\ for the bremsstrahlung.

\section{The 1.8~\MakeLowercase{h} pulsator: \cv}\label{cataclysmic}

\subsection{Timing analysis}

\begin{figure*}
\centering
\resizebox{\hsize}{!}{\includegraphics[angle=-90]{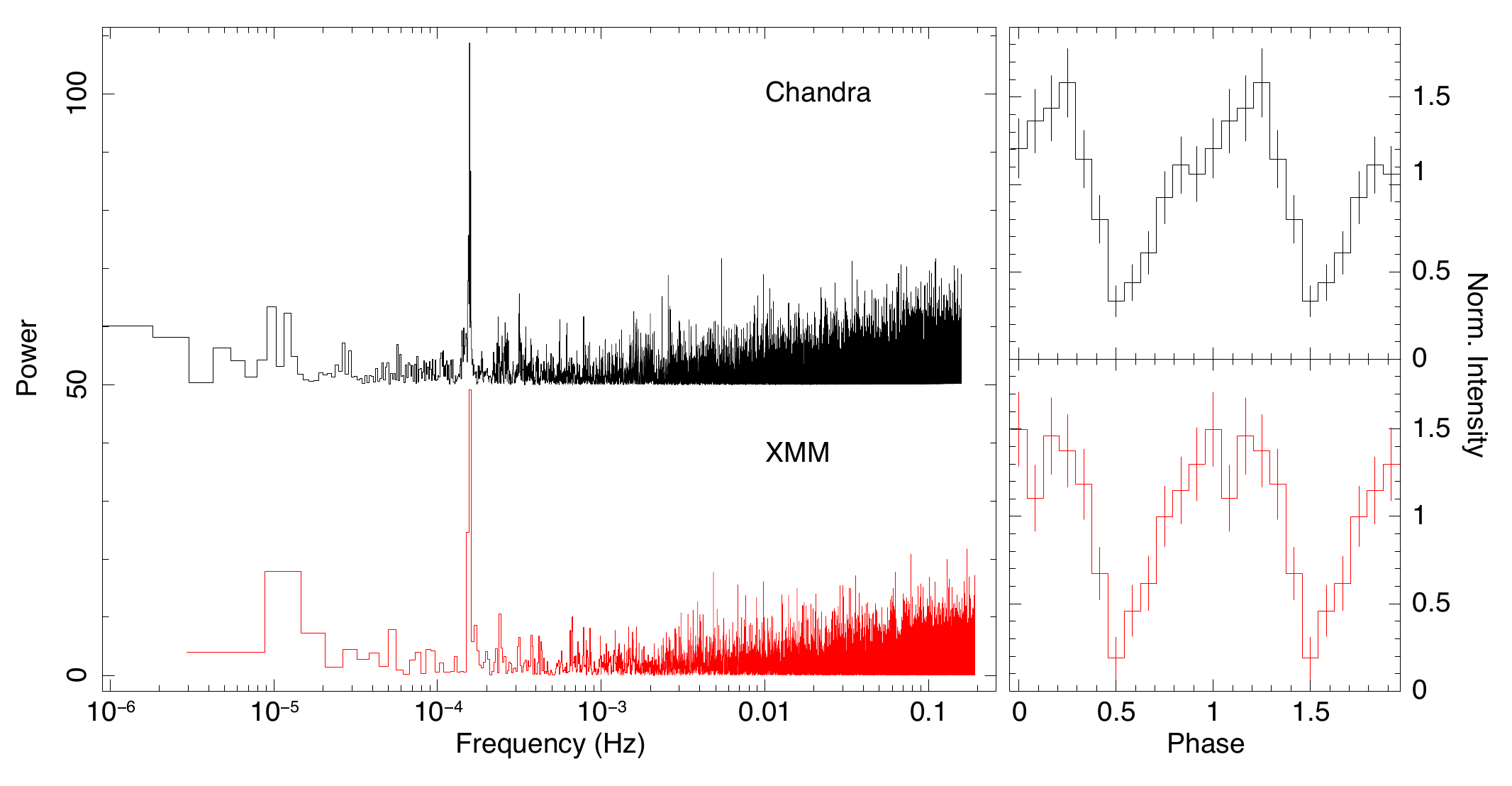}}
\caption{\label{dpsfold1413} Same as Figure \ref{dpsfold1414} but for \cv. For displaying purposes, the \cxo\ data have been shifted in power by +50.}
\end{figure*}

\cv\ displays a periodicity at roughly 6.4~ks (Fig.\,\ref{dpsfold1413}). This signal was detected at about 3.5$\sigma$ confidence level in observation 12823 ($\sim$270 counts) and is present also in the shorter pointing 12824 ($\sim$80 counts), though at a lower confidence level. 
By the phase-fitting analysis, we derived the value $P=6\,377\pm4$~s. In the corresponding 0.5--10 keV background-subtracted folded profile (Fig.\,\ref{dpsfold1413}), we measured a pulsed fraction of $56\pm8$~per~cent. The modulation of \cv\ changes in shape and pulsed fraction as a function of energy: the asymmetry in the profile becomes more accentuated, while the pulsed fraction decreases as the energy increases: $78\pm7$~per~cent in the soft range ($<$2~keV) and $40\pm8$~per~cent in the hard range ($>$2~keV). Of all the other observations in Table\,\ref{obslog}, \cv\ was detected only in the 2001 \xmm\ data, where the harvest was $\sim$440 source photons between the pn and MOS\,1. We measured a period of $6.4\pm0.1$~ks and the pulsed fraction was $59\pm10$~per~cent. 

\subsection{Spectral analysis}

Given the paucity of photons (all in the 1--6~keV range), we performed the spectral analysis by fitting the same spectral models adopted for \ip\ to the two \cxo\ data sets 12823 and 12824 simultaneously, with the normalisations free to vary and the other parameters tied up. The blackbody model provides a poor fit to the data, with a $\chi^2_\nu=1.68$ for 13 dof, while for the bremsstrahlung ($\chi^2_\nu=1.31$) the temperature could not be constrained. For this reasons, in Table\,\ref{cv_specs} we give only the parameters obtained from the power law fit  (see Fig.\,\ref{cvs_spec}, right), which gave a good fit with a rather hard power law with photon index $\Gamma\approx0.8$ and a best-fitting absorption $N_{\rm H}=0.2\times10^{22}$~cm$^{-2}$ (with a 90~per~cent upper limit of $0.9\times10^{22}$~cm$^{-2}$). The observed flux was $F_{\mathrm{X}}\approx4\times10^{-14}$~\flux\ during the first observation, and $\approx$$6\times10^{-14}$~\flux\ in the second; when one considers the large uncertainties, the flux increase is however only marginally significant.
\begin{table*}
\begin{minipage}{15.5cm}
\centering \caption{Spectral results of \cv. Errors are at a 1$\sigma$ confidence level for a single parameter of interest.} \label{cv_specs}
\begin{tabular}{@{}lccccccc}
\hline
Model & Data & \nh$^a$ & $\Gamma$ & $kT$ & Flux$^b$ & Unabsorbed flux$^b$ & $\chi^2_\nu$ (dof) \\
 & & ($10^{22}$~cm$^{-2}$) &  & (keV) & \multicolumn{2}{c}{($10^{-14}$ \flux)} &  \\
\hline
\multirow{2}{*}{\textsc{phabs(powerlaw)} }& \cxo/12823 & \multirow{2}{*}{$<$$0.9^c$} & \multirow{2}{*}{$0.8^{+0.3}_{-0.2}$} & \multirow{2}{*}{--} & $4.3^{+0.8}_{-0.6}$ & $4.5^{+0.6}_{-0.5}$ & \multirow{2}{*}{1.02 (13)}\\
 & \cxo/12824 &  &  & & $6.3^{+1.4}_{-1.3}$ & $6.5^{+1.2}_{-1.0}$ & \\
\hline
\textsc{phabs(powerlaw)} & \emph{XMM}/0111240101 & $0.13^{+0.09}_{-0.07}$ & $0.93\pm0.13$ & -- & $5.4\pm0.5$ & $5.6^{+0.5}_{-0.4}$ & 0.82 (16)\\
\hline
\end{tabular}
\begin{list}{}{}
\item[$^{a}$] The abundances used are those of \citet*{wilms00}; \nh\ values $\approx$30~per~cent lower are derived with those by \citet{anders89}. The photoelectric absorption cross-sections are from \citet{balucinska92}.
\item[$^{b}$] In the 0.5--10~keV energy range. 
\item[$^{c}$] Upper limit at 90~per~cent confidence level. 
\end{list}
\end{minipage}
\end{table*}

The 2001 \xmm\ observation offers a better spectrum, covering with a few more photons the band 0.4--8~keV. The power-law fit ($\chi^2_\nu=1.02$ for 16 dof) yielded photon index and flux similar to those derived with \cxo\ (see Table\,\ref{cv_specs}) and made it possible to constrain better the absorbing column. This was measured at $N_{\rm H}=0.13^{+0.09}_{-0.07}\times10^{22}$~cm$^{-2}$. 
In the 2014 \xmm\ observation, \cv\ was not detected. The 3$\sigma$ upper limit on its observed flux, derived with \textsc{pimms} assuming the \xmm\ power law model in Table\,\ref{cv_specs}, was $2\times10^{-14}$~\flux\ in the 0.5--10~keV band in the MOS data (in the pn the position of the source occurred in proximity to streaks of out-of-time events due to the nucleus of the CG). 

The position of \cv\ was imaged also in the \cxo\ observations 2454, 355, 356, 10850, 10872, and 10873 (see Table\,\ref{obslog}); the source was never detected and for each data set we derived in a like manner the following upper limits (for the grating observations, we considered only the zero-order data): $1.1\times10^{-13}$~\flux\ (obs. 355, ACIS-I), $10^{-14}$~\flux\ (obs. 356, ACIS-S), $7\times10^{-14}$~\flux\ (obs. 2454, ACIS-S), $5\times10^{-14}$~\flux\ (obs. 10873, ACIS-S), $8\times10^{-14}$~\flux\ (obs. 10850, ACIS-S), and $6\times10^{-14}$~\flux\ (obs. 10872, ACIS-S).

\section{Astrometry and optical observations of \ip\ and \cv}\label{vst_data}
\subsection{X-ray astrometry}

In order to improve the absolute astrometry of the \cxo\ data to search for optical counterparts to \ip\ and \cv, we cross-correlated the X-ray source list obtained using \textsc{wavedetect} with sources in the Two-Micron All-Sky Survey (2MASS; \citealt{skrutskie06}) catalog, which has an astrometric accuracy better than $0\farcs2$.
We found 17 2MASS point sources coincident within $0\farcs4$ from an X-ray source and used them to register the \cxo\ images on the accurate 2MASS reference frame by fitting a transformation matrix which includes a rotation, scale factor, and translation. We note that the \cxo--2MASS superposition did not require a significant transformation: the corrections are of the same order of the residuals ($<$$0\farcs15$). The resulting positions (J2000.0) of the new CATS\,@\,BAR pulsators are $\rm RA = 14^h14^m30\fs1$ ($\pm$$0\fs45$) and $\rm Decl. = -65\degr16'23\farcs3$ ($\pm$$0\farcs30$) for \ip, and $\rm RA = 14^h13^m32\fs9$ ($\pm0\fs30$) and $\rm Decl. = -65\degr17'56\farcs5$ ($\pm$$0\farcs25$) for \cv, where the 1$\sigma$ uncertainties combine the \cxo\ localisation accuracy, the residuals of the \cxo--2MASS frame superposition, and the 2MASS absolute astrometric accuracy.

\subsection{VST data}

Optical images of the regions around \ip\ and \cv\ were retrieved from the ESO Science Archive Facility. The observations were originally obtained with the 2.6-m VST located at Paranal Observatory using the OmegaCAM instrument \citep{kuijken11}, as part of the VST Photometric H$\alpha$ Survey of the Southern Galactic Plane and Bulge (VPHAS+; \citealt{drew14}).  Table\,\ref{tab:VSTimages} lists the details of the images examined.
\begin{table}
\centering
\caption{VST/OmegaCAM images of regions around \ip\ and \cv. \label{tab:VSTimages}}
\begin{tabular}{@{}lccc}
\hline
MJD    & Exp.&       Filter &           Archive name\\
 & (s) & & \\
           \hline
56442.149	& 25 & $r'$ & OMEGA.2013-05-30T03:34:12.715\\
56442.156	& 25 & $r'$ & OMEGA.2013-05-30T03:44:47.517\\
56460.067	& 25 & $r'$ & OMEGA.2013-06-17T01:07:16.243\\
56460.054	& 25 & $r'$ & OMEGA.2013-06-17T01:18:02.485\\
56487.025	& 25 & $r'$ & OMEGA.2013-07-14T00:36:46.111\\
56487.014	& 40 & $r'$ & OMEGA.2013-07-14T00:20:12.159\\
56488.033	& 25 & $r'$ & OMEGA.2013-07-15T00:47:30.554\\
\hline
\end{tabular}
\end{table}

We aligned and stacked the images using the Image Reduction and Analysis Facility (\textsc{iraf}) software and the \textsc{imalign} and \textsc{imcombine} packages following standard procedures. A world coordinate system was then applied using the \textsc{imwcs} utility\footnote{See \mbox{http://tdc-www.harvard.edu/wcstools/}.} and the Third US Naval Observatory CCD Astrograph Catalog (UCAC3; \citealt{zacharias10}). The mean error in the world coordinate system is $0\farcs 35$ (3$\sigma$) using 22 local stars. We performed aperture photometry on the nearby fields using the \textsc{iraf} package \textsc{apphot} and estimate a limiting magnitude depth of $r' \approx22.5$ mag. No sources are observed within $2\arcsec$ of the coordinates of \ip\ and \cv.

\section{Discussion~I: the nature of \ip\ and \cv}\label{discussion}

With a new X-ray pulsator discovered by CATS\,@\,BAR every $\sim$2,200 light curves, the detection of two previously unknown pulsating sources in the Circinus data set had a formal probability of about 0.04~per~cent. While this does not qualify as a statistical anomaly, the 2010 \cxo\ observations have certainly been bountiful for the CATS\,@\,BAR project.

\subsection{\ip}
In the case of \ip, which is displaced by $\sim$2~arcmin from the extreme edge of CG (Fig.\,\ref{circinus_ds9}), the two periodicities nail it down as an intermediate polar (IP) with spin period of $P_{\mathrm{spin}}=6.1$~ks (1.7~h) and orbital period $P_{\mathrm{orb}}=64.2$~ks (17.8~h). IPs, also known as DQ~Herculis stars, and polars (AM Herculis stars) are the two main subclasses of magnetic cataclysmic variable stars (CVs). CVs are close binaries hosting a white dwarf (WD) accreting from a late-type Roche-lobe filling companion, either a main-sequence or a sub-giant star. IPs are characterised by asynchronous rotation ($P_{\mathrm{spin}}<P_{\mathrm{orb}}$), while polars are phase-locked ($P_{\mathrm{spin}}\simeq P_{\mathrm{orb}}$) and generally display strong circular polarisation (whence the name) at optical and near infrared wavelengths (see \citealt{patterson94,warner03,smith06} for reviews). Although it is still matter of debate, these differences are generally interpreted as due to a magnetic field in IPs which is weaker than that typically measured for polars ($B\ga10^7$~G).

The orbital period of \ip\ locates the system above the 2--3~h so-called `orbital period gap', where most of the IPs are found. Also the spin-to-orbit period ratio of $\sim$0.095 is typical of an IP (it is generally in the range 0.25--0.01, with most systems around 0.1).\footnote{See for example the catalog available at the Intermediate Polar Home Page, \mbox{http://asd.gsfc.nasa.gov/Koji.Mukai/iphome/catalog/members.html}.} In IPs, the accreted material generally passes through a disc and is then channelled onto the magnetic polar regions of the WD (at variance with polars, where the magnetic field inhibits the formation of the disc). There, a shock develops and the hot gas cools while it settles onto the WD surface emitting X-rays via thermal bremsstrahlung and cyclotron radiation \citep{aizu73}. Because of their strong magnetic field, in polars the cooling takes place mainly via cyclotron, whereas IPs are expected to show bremsstrahlung-dominated emission. While the relatively poor statistical quality of the available spectra precluded a good characterisation of the X-ray emission of \ip, the results of the spectral analysis are consistent with this picture (Table\,\ref{ip_specs}). Also the large-amplitude modulation at the orbital period is rather common in IPs (e.g. \citealt*{parker05}), and the $\sim$100~per~cent pulsed fraction hints a high-inclination system. We finally note that the X-ray luminosity of \ip\ in the deep \cxo\ and \xmm\ observations was $L_{\mathrm{X}}\approx2\times10^{31}d^2_1$~\lum\ (with a $\approx$50~per~cent variability, see Section\,\ref{polar}), where $d_1$ is the distance in units of $1$~kpc. Typical values for IPs, $10^{32}$--$10^{34}$~\lum\ \citep{sazonov06}, suggest that \ip\ is either on the lower side of the luminosity distribution or the distance to the source is substantially larger than $\sim$1~kpc.  

For an IP with orbital period of 17.8~h, a K5V star would be a likely companion (e.g. \citealt{smith98}). Using a value of $N_{\rm H} = 0.3 \times 10^{22}$~cm$^{-2}$ derived from our model fit to the X-ray spectra, and a conversion of $N_{\rm H}/A_V$ of $1.79 \times 10^{21}$~cm$^{-2}$~mag$^{-1}$ \citep{predehl95}, we obtain $A_V = 1.7$ mag. Assuming an absolute magnitude $M_{r'} \approx 7.1$ mag \citep{bilir08},\footnote{Uncertainty in the assumed absolute magnitudes of the companion star may be as large as 2~mag. The same holds for \cv, see below.} the limiting magnitude $m_{r'} \approx 22.5$ suggests $d > 5$~kpc.

\subsection{\cv}
The nature of the fainter \cv\ is less obvious. The source is located inside the 90~per~cent total $B$ light contour of the CG. So, the first question that needs to be addressed is whether it is a Galactic or an extragalactic source. Based on the cumulative Galactic X-ray source density versus flux distribution ($\log N$--$\log S$) from the \cxo\ Multi-wavelength Plane survey (ChaMPlane; \citealt{vandenberg12}), we estimated the probability of a foreground Galactic object of that flux within this area to be $\approx$11~per~cent. Moreover, its absorbing column measured with \xmm\ ($N_{\rm H}=1.3^{+0.9}_{-0.7}\times10^{21}$~cm$^{-2}$; Table\,\ref{cv_specs}) is much lower than the total Galactic value of $\sim$$6\times10^{21}$~cm$^{-2}$ \citep{dickey90,kalberla05}. \cv\ is therefore most likely a Galactic source. 

The hard X-ray spectrum, a power law with photon index $\Gamma\sim0.8$--0.9, and the modulation at 6.4~ks (1.8~h) point to a binary system consisting of a compact star accreting from a low-mass companion, where the 1.8~h period likely traces the orbital motion. The period is in fact too long to be the spin of a typical neutron star (NS; with very few possible exceptions; \citealt{mattana06,etdl11}). The low flux and---chiefly---the smooth and $\sim$60~per~cent pulsed-fraction modulation favour a magnetic CV nature also for \cv\ (the period is also too short for the orbit of a standard high-mass X-ray binary). Indeed, CVs are the most abundant population of Galactic compact interacting binaries, and also the most frequent new pulsating sources in the CATS\,@\,BAR sample (Israel et al., in preparation). In particular, since no second periodicity was detected (and lacking any information about optical polarisation), the source could be a polar. Polars are generally found at short orbital periods, most of them below the 2--3~h orbital gap, and \cv\ would lie in the peak of their period distribution (e.g. \citealt{ritter03}). The profile of the folded light curve (Fig.\,\ref{dpsfold1413}) and its variability as function of energy may indicate a two-pole system.
The luminosity during the long 2010 \cxo\ observations was  $L_{\mathrm{X}}\approx(5$--$8)\times10^{30}d^2_1$~\lum, with the upper limits from the other observations implying a variability of $\approx$50~per~cent or larger. For distances of the order of a few kpc, this is in good agreement with typical values for polars ($L_{\mathrm{X}}<10^{32}$~\lum; \citealt{sazonov06}).

For a polar with orbital period of 1.8~h, the companion is likely a M5V star, which has an absolute magnitude of $M_{r'} \approx 12.5$ mag \citep{bochanski11}. The value of $N_{\rm H} = 0.13 \times 10^{22}$~cm$^{-2}$ derived from our model fit to the X-ray spectra implies an $A_V = 0.7$ mag. The limiting magnitude $m_{r'} \approx 22.5$ suggests $d \ga 0.7$~kpc. In the IP hypothesis, assuming for \cv\ a K5V as for \ip, the non-detection would indicate a distance larger than $\sim$8~kpc.

\section{The controversial source CG\,X--1}\label{cgx1}

\cgx\ (CXOU\,J141312.3--652013), about 15~arcsec northeast of the Circinus' nucleus, had been known for long to be a bright and variable (possibly periodic) X-ray source. Using high-quality light curves collected with \cxo, \citet{bauer01} discovered a strong modulation at a period of $\sim$27~ks. The measured X-ray flux was $9\times10^{-13}$~\flux\ (0.5--10~keV), and deep \hst\ observations did not detect any optical counterpart to \cgx, with a limit $m_{F606W}>25.3$. They observed that the source might be either a black hole (BH) binary in the CG radiating at $\sim$$4\times10^{39}$~\lum\ (and hence qualifying as an ULX) or a Galactic cataclysmic variable of the polar type with a particularly long period (in both cases, the 27-ks modulation would reflect the orbital period of the system). 

In the polar hypothesis, for a M2V to M6V companion star, the \hst\ limit puts the source at a distance larger than 1.2~kpc, implying a luminosity of at least $3\times10^{32}$~\lum, a rather extreme value for a polar. \citet{bauer01} also noticed that the association of \cgx\ with the CG is convincing: based mainly on the results from the \emph{ASCA} Galactic plane survey \citep{sugizaki01}, they evaluated that the possibility of a foreground or background X-ray source is $\lesssim$0.06~per~cent. Moreover, \citet{sw01} noticed that the absorption toward \cgx\ ($N_{\rm H}>10^{22}$~cm$^{-2}$) is much larger than the total Galactic column ($N_{\rm H}\sim6\times10^{21}$~cm$^{-2}$), further supporting the association with the CG. Overall, \citet{bauer01}, \citet{sw01}, and \citet{bianchi02} favoured a very bright extragalactic BH binary, harbouring a BH possibly in excess of $50$~M$_{\sun}$.

Despite recognising the robustness of the association, \citet{weisskopf04} were more open toward the possibility of a foreground polar. They observed that while the period and the luminosity would be somewhat atypical, \cgx\ would be neither the longest-period, nor the brightest known polar. On the other hand, they argued that if \cgx\ belonged to the CG, because of the short orbital period it should be a BH low-mass X-ray binary (LMXB) with a $\lesssim$1~M$_{\sun}$ companion. In this case, the huge X-ray luminosity of the BH would drive the star out of thermal equilibrium and evaporate it within $\sim$$10^3$~yr. They regard as very unlikely the possibility that a system this short-lived could be observed.

\citet{weisskopf04} revised the period of \cgx\ at $26.25\pm0.15$~ks and noticed a possible optical counterpart with $m_{F606W}=23.5$. However, they did not estimate the significance of the association or of the source, and gave instead a limiting magnitude of 24.3. \citet{ptak06} confirmed the limit by  \citet{bauer01}, while a recent work by \citet{gladstone13} proposed a counterpart with $m_V=24\pm6$ (presumably, the same excess/source detected by \citealt{weisskopf04}). In literature, \cgx\ is generally considered to be an ULX in the CG (e.g. \citealt{bianchi02,swartz04,lm05,ptak06,berghea08,gladstone13}).

\section{The new Chandra data of \cgx: Analysis and results}\label{gcx_analysis}

Since the CG has been observed many times in X-rays, in particular with \cxo, a wealth of data exist for \cgx. Detailed studies of \cgx\ with \rst, \sax, \xmm, and \cxo\ were presented in the aforementioned works by \citet{sw01}, \citet{bauer01}, \citet{bianchi02}, and \citet{weisskopf04} (but see also \citealt{matt96,guainazzi99,sambruna01,massaro06,bauer08,yang09,shu11,walton13,arevalo14}). A systematic analysis of all the available data is beyond the scope of this paper, and here we will only present results from the analysis of observations 12823/4, which represent the deepest and highest-quality data set available for this source and, to our knowledge, have not been used so far to study \cgx.

The coordinates we derived for \cgx\ are $\rm RA = 14^h13^m12\fs24$ ($\pm0\fs25$)  and $\rm Decl. = -65\degr20'13\farcs82$ ($\pm$$0\farcs25$). Using recent results from the ChaMPlane survey \citep{vandenberg12}, we confirm the low probability of a foreground or background object ($\lesssim$0.06~per~cent) estimated by \citet{bauer01}. We actually believe that the probability of an interloper is substantially lower (approximately two times smaller), considering that the flux reached by the source in subsequent observations is $\sim$5 times higher \citep{weisskopf04} and the fact that a background AGN can be excluded by the phenomenology of \cgx.

For the spectral and timing analysis, we extracted the source counts within a 1.5-arcsec radius, while for the background we used an annulus with radii of 3 and 5 arcsec (see Fig.\,\ref{circinus_ds9} and Section\,\ref{observations} for more details). The 0.3--8~keV source net count rate was $(6.90\pm0.07)\times10^{-2}$~counts~s$^{-1}$ in observation 12823, and $(6.1\pm0.1)\times10^{-2}$~counts~s$^{-1}$ in observation 12824. These rates are high enough to cause pileup in the ACIS detector (as we checked with a pileup map). For the spectral analysis, the pileup was dealt with by using the pileup model by \citet{davis01}. This procedure involves some uncertainty, because the pileup fraction in \cgx\ is strongly dependent on the orbital phase. However none of our results crucially depends on the exact value of the parameters derived from the spectral fitting. 

\begin{figure}
\centering
\resizebox{\hsize}{!}{\includegraphics[angle=-90]{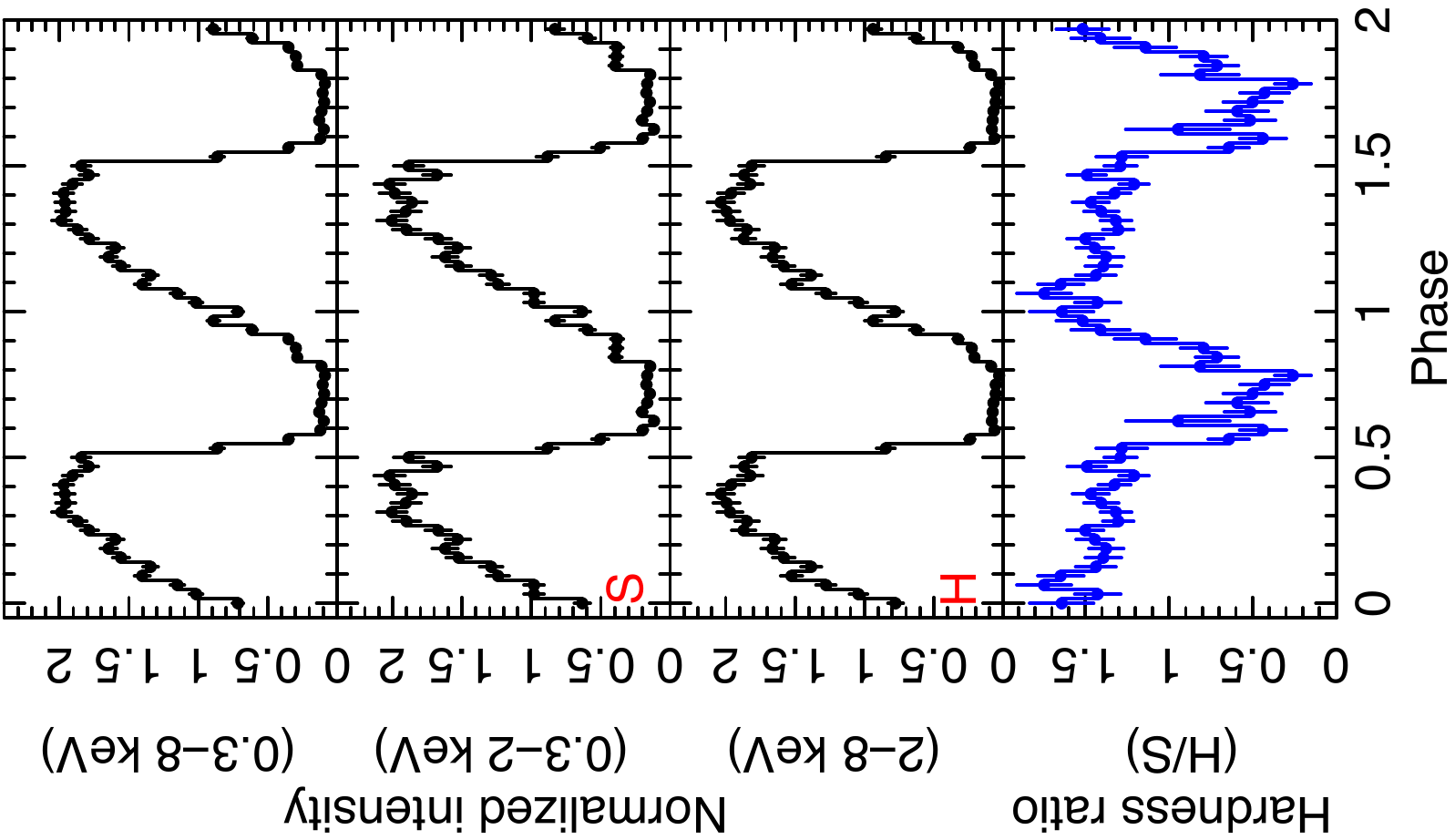}}
\caption{\label{cgx1_efold} Background-subtracted folded profile of CG\,X--1 (observation 12823/4) in different energy bands. The hardness ratio between the hard and soft bands is also plotted at the bottom.}
\end{figure}

About six 26-ks cycles were recorded in obs. 12823 and two in obs. 12824. There is a moderate pulse-to-pulse variability, both in shape and in the flux at maximum ($\approx$30~per~cent). Also, the average count rate was $\approx$15~per~cent lower during the second observation. To measure the period, we fit a sinusoidal function to the light curve. We obtained $P_{\mathrm{CG\,X\text{--}1}}=26.1\pm0.1$~ks. The corresponding folded profile in different energy bands is shown in Fig.\,\ref{cgx1_efold}.
The profile is asymmetric and the modulations is large (but the count rate is non-zero also at minimum). The pulsed fraction, defined as\footnote{Here we adopted a different definition of the pulsed fraction than used before, because the value inferred from a sinusoidal fit would misrepresent the amplitude of the modulation of the `sawtooth' profile of \cgx.} $(M-m)/(M+m)$, where $M$ is the maximum count rate and $m$ the minimum, is $91.6\pm1.5$~per~cent in the whole band, $89.2\pm2.5$~per~cent in the 0.3--2~keV band, and $97.1\pm1.2$~per~cent in the 2--8~keV band. A conspicuous spectral softening around minimum is evident from the ratio of the hard and soft counts along the cycle (Fig.\,\ref{cgx1_efold}).

\begin{figure*}
\centering
\resizebox{0.7\hsize}{!}{\includegraphics[angle=-90]{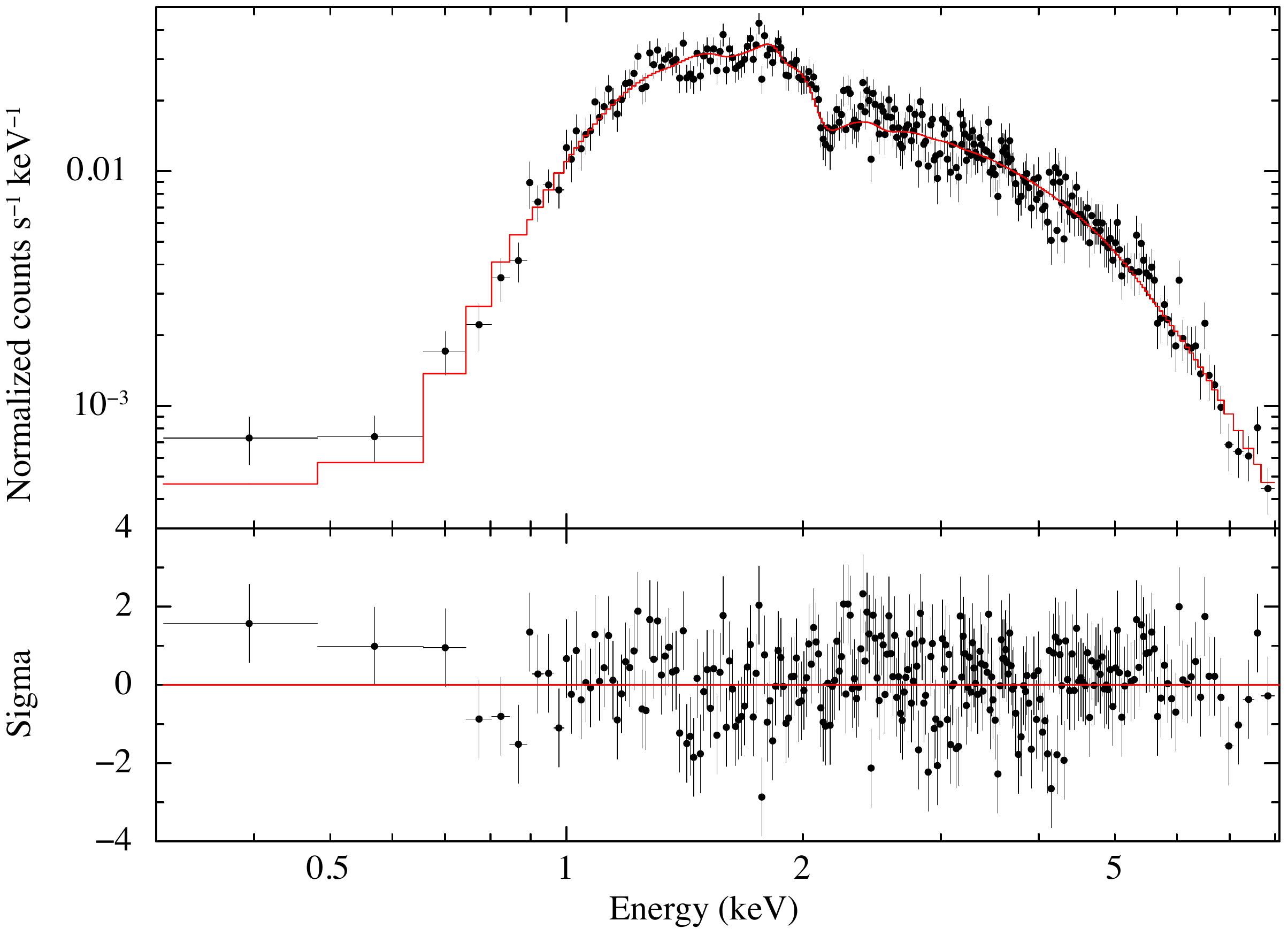}}
\caption{\label{cgx1spec} \cxo/ACIS spectrum and best-fitting MCD model (red solid line) for \cgx\ from observation 12823. Bottom panel: the residuals of the fit in units of standard deviations.}
\end{figure*}

For the spectral analysis we fit three simple models to the data: a power law, a multicolour disc (MCD; \citealt{mitsuda84,makishima00}), and an optically-thin thermal bremsstrahlung, all corrected for the interstellar absorption. While a multicolour disk with $kT\simeq1.3$--1.4~keV gives the lowest $\chi^2$, all models provide an acceptable fit to the data (Table\,\ref{cgx1_specs}). All fits confirm that \nh\ towards \cgx\ is substantially larger than the total Galactic absorbing column in that direction. \citet{weisskopf04} reported a possible feature, probably a blend of Fe lines, in the 2001 \xmm\ spectrum of \cgx. No line is required to fit the \cxo\ data (see Fig.\,\ref{cgx1spec} for the longest observation); the 3$\sigma$ upper limit on the equivalent width of any line with central energy between 6 and 7~keV is 0.18~keV in observation 12823. This limit is formally compatible with the equivalent width of $0.23\pm0.06$~keV derived by \citet{weisskopf04}. However, since there is no trace of such feature in the ACIS data, it is possible that, as noticed also by \citet{weisskopf04}, the feature observed with \xmm\ is due to residual contamination from the emission of the nuclear region of the CG, which has very strong lines at 6.4 and 7~keV. 

We also extracted pulse-resolved spectra from the soft (phase between 0.55 and 0.9 in Fig.\,\ref{cgx1_efold}) and hard (all other phases) parts of the hardness ratio. The data were prepared with \textsc{dmtcalc}, and the spectra of the two data sets from the same phase bins were combined using \textsc{combine\_spectra}, which also averaged the response matrices. Although the few counts in the soft spectrum ($\sim$700 versus more than 12\,200 in the hard spectrum, after combining the two observations) preclude a detailed comparison, we found that the spectral variation can be described equally well by either a decrease in the absorption during the softening ($N_{\rm H}=(1.00\pm0.03)\times10^{22}$~cm$^{-2}$ in the hard phase and $(0.24\pm0.07)\times10^{22}$~cm$^{-2}$ in the soft phase for the MCD model; $(1.45\pm0.04)\times10^{22}$~cm$^{-2}$ in the hard phase and $(0.67\pm0.08)\times10^{22}$~cm$^{-2}$ in the soft phase for the power-law model; $(1.29\pm0.03)\times10^{22}$~cm$^{-2}$ in the hard phase and $(0.51\pm0.07)\times10^{22}$~cm$^{-2}$ in the soft phase for the bremsstrahlung model) or a change in the pivotal parameter of the model ($kT=1.82\pm0.05$~keV in the hard phase and $0.95\pm0.06$~keV in the soft phase for the MCD; $\Gamma=1.70\pm0.03$ in the hard phase and $2.54\pm0.11$ in the soft phase for the power law; $kT = 9.4\pm0.8$~keV in the hard phase and $2.4\pm0.3$~keV in the soft phase for the bremsstrahlung).
\begin{table*}
\begin{minipage}{16.15cm}
\centering \caption{Spectral results of CG\,X--1. Errors are at a 1$\sigma$ confidence level for a single parameter of interest.} \label{cgx1_specs}
\begin{tabular}{@{}lccccccc}
\hline
Model & Obs.\,ID & \nh$^a$ & $\Gamma$ & $kT$ & Flux$^b$ & Unabsorbed flux$^b$ & $\chi^2_\nu$ (dof) \\
 & & ($10^{22}$ cm$^{-2}$) &  & (keV) & \multicolumn{2}{c}{($10^{-12}$ \flux)} &  \\
\hline
\textsc{phabs(diskbb)} & 12823 & $1.06\pm0.04$ &  --  & $1.36\pm0.07$ & $0.83^{+0.13}_{-0.12}$ & $1.17\pm0.17$ & 0.97 (271)\\
\textsc{phabs(powerlaw)} & 12823 & $1.38\pm0.05$ & $1.72_{-0.05}^{+0.06}$ & -- & $1.23^{+0.10}_{-0.09}$ & $1.84^{+0.12}_{-0.10}$ & 1.04 (271)\\
\textsc{phabs(bremsstrahlung)} & 12823 & $1.28\pm0.04$ &  --  & $6.5^{+1.0}_{-0.8}$ & $1.02^{+0.04}_{-0.05}$ & $1.50^{+0.05}_{-0.06}$ & 1.00 (271)\\
\hline
\textsc{phabs(diskbb)} & 12824 & $1.03^{+0.08}_{-0.07}$ &  --  & $1.33^{+0.14}_{-0.10}$ & $0.68^{+0.21}_{-0.16}$ & $0.85^{+0.29}_{-0.22}$ & 0.98 (92)\\
\textsc{phabs(powerlaw)} & 12824 & $1.38^{+0.11}_{-0.10}$ & $1.77_{-0.10}^{+0.12}$ & -- & $1.00^{+0.15}_{-0.14}$ & $1.52^{+0.18}_{-0.17}$ & 1.11 (92)\\
\textsc{phabs(bremsstrahlung)} & 12824 & $1.27\pm0.08$ &  --  & $5.6^{+1.7}_{-1.2}$ & $0.83\pm0.07$ & $1.25\pm0.09$ & 1.05 (92)\\
\hline
\end{tabular}
\begin{list}{}{}
\item[$^{a}$] The abundances used are those of \citet*{wilms00}; \nh\ values $\approx$30~per~cent lower are derived with those by \citet{anders89}. The photoelectric absorption cross-sections are from \citet{balucinska92}.
\item[$^{b}$] In the 0.5--10~keV energy range. 
\end{list}
\end{minipage}
\end{table*}

\section{Discussion~II: Is \cgx\ a Wolf--Rayet/black-hole binary in the Circinus galaxy?}\label{cgx_discussion}

While we cannot completely exclude the chance superposition of a Galactic polar in the direction of the inner part of the CG, we regard the association of the \cgx\ to the CG as rather compelling. In the following, we will discuss the nature of the source in this framework.

\citet{weisskopf04} disfavoured the possibility of a BH LMXB because such a system would be rather short lived, and thus unlikely to be observed. This argument is not conclusive when dealing with an individual source (one can be lucky enough to observe a rare system!), but we too deem a LMXB as unlikely. This is because of its orbital profile (Fig.\,\ref{cgx1_efold}). In fact, most LMXBs show no or very low-amplitude modulation on their orbital period. In those which display orbital modulation, dipping and/or eclipsing systems, the morphology of the profile is very different. In dipping systems, the dips are produced by absorption of X-rays due to accreting matter located in the bulge at the outer edge of the accretion disc (e.g. \citealt{diaztrigo06}); the X-ray minima tend to be rather sharp and to show a harder than average emission, which is the opposite of what is observed in \cgx\ (Fig.\,\ref{cgx1_efold}, bottom panel). LMXB eclipses display sharp and abrupt ingresses and egresses, due to the small size of the X-ray-emitting regions. Even the smoothly-modulated accretion disc corona sources (systems viewed nearly edge-on where the outer edge of the dense disc modulates the X-rays from the central source scattered into the line of sight by an extended ionised corona; \citealt{mason82,white82,somero12}) have different profiles. Moreover, LMXBs containing BHs are usually transient X-ray sources, with outburst durations of the order of weeks to months (e.g. \citealt{remillard06}), while \cgx\ is variable but persistent.

On the other hand, the light curve and the folded profile of \cgx\ bear a strong resemblance to those observed in high-mass X-ray binaries (HMXBs) with a Wolf--Rayet (WR) star companion: Cyg\,X--3 in the Milky Way ($P=4.8$~h; \citealt{zdziarski12}), IC\,10~X--1 in IC\,10 ($P=35$~h; \citealt{prestwich07}), NGC\,300~X--1 in NGC\,300 ($P=33$~h; \citealt{cpp07}), and the candidates CXOU\,J123030.3+413853 in NGC\,4490 (CXOU\,J123030; $P=6.4$~h; \citealt{eism13}) and CXOU\,J004732.0--251722.1 in NGC\,253 (with a candidate periodicity $P\sim14$--15~h; \citealt{maccarone14}).\footnote{Apart from these objects, the only other known WR HMXB is ULX--1 in M101 (\citealt{liu13}, see also Section\,\ref{demographics}). A period of 8.2\,d was inferred from radial velocities of optical emission lines, but no X-ray light curves of good quality are available for this source.} 
The large modulation of these systems, showing a slow rise and a faster decay, is different from what observed in almost any other class of X-ray sources (see, e.g., the arguments summarised in \citealt{eism13} about the source in NGC\,4490), and can be considered the signature of WR HMXBs.

Although its properties remain still poorly understood, the best studied source of this kind is Cyg\,X--3, where the orbital modulation has been ascribed to attenuation by electron scattering in the strong WR wind \citep{hertz78}. A further significant contribution to the scattering medium is thought to be a bulge of ionised matter formed by the collision of the stellar wind with the outer accretion disc \citep{zdziarski10}. Alternative explanations involve orbital modulation by absorbers in different phases (hot/ionised and cold/clumpy, triggered by the BH jet bow shock;  \citealt{vilhu13}), or the presence of the accretion wake, a large scale asymmetry around the compact object \citep{okazaki14}. The orbital profile of \cgx\ is even more asymmetric than in Cyg\,X--3, probably due to rather extreme properties of the absorbing/scattering medium or because of a higher inclination of the system. 
The variation of the hardness ratio along the orbit (Fig.~\ref{cgx1_efold}), showing a clear softening during the X-ray minimum, could be due to direct X-rays almost completely blocked by dense matter (probably the innermost regions of the WR wind when the X-ray source is at the superior conjunction). The softer X-rays observed could be due to down-scattering into the line of sight of central X-rays by cold and less dense material located farther away from the WR star (similarly to what usually observed during eclipses in some HMXBs; \citealt{haberl91}).

\citet{weisskopf04} mentioned for \cgx\ the possibility of a WR or, more in general, a naked He donor, but did not discuss it in detail owing to the scarcity of information of such systems. In particular, they observed that for a $\sim$2~M$_\odot$ companion, their argument against LMXB systems was still relevant. However, thanks to their compactness, even more massive WR stars can comfortably fit in the orbit of a system like \cgx\ (the stellar radius of a 20-M$_\odot$ WR star is $<$2~R$_\odot$; e.g. \citealt{langer89,schraerer92}). Also the other main observational properties of \cgx\ fit well in the scenario of a WR--BH HMXB in the CG.\footnote{Even bright ($L_{\mathrm{X}}>10^{40}$~\lum) ULXs may contain a NS  \citep{king09,bachetti14}, although such systems are probably not the majority \citep{fragos15}. Here, we will not discuss this possibility.} \\

If one considers the reddening toward the CG (4~mag) and its distance module (28.1~mag, from NED), the limit on the optical counterpart by \citet{bauer01} and \citet{ptak06} implies $M_V>-6.8$. This value is compatible with a WR star, for which $M_V$ is typically in the range from $-2.5$ to $-7$ (e.g., \citealt{massey03}). 

The X-ray luminosity of \cgx\ is variable by a factor of $\approx$10 \citep{bianchi02,weisskopf04}. The highest flux reported in literature ($5.2\times10^{-12}$~\flux\ for a power-law fit or $5\times10^{-12}$~\flux\ for a MCD fit, in the 0-5--8~keV band; \citealt{weisskopf04}) would imply, for a distance of 4.2~Mpc, a 0.5--10~keV luminosity of $L_{\mathrm{X}}=(1.5$--$2)\times10^{40}$~\lum. 
If the system is Eddington-limited, the lower limit on the mass of the accreting BH is $M_{\mathrm{BH}}\gtrsim75$~M$_\odot$ for a He or C/O donor.
For the system to shine in X-rays, the velocity of the WR star wind has to be slow enough to allow the formation of an accretion disk. This condition corresponds for \cgx\ to the requirement $M_{\mathrm{BH}}\ga1.5\,v_{\mathrm{w,\,1000}}^4\delta^2$~M$_\odot$, where $M_{\mathrm{BH}}$ is the BH mass, $v_{\mathrm{w,\,1000}}$ is the wind velocity in units of 1000~km~s$^{-1}$, and $\delta\approx1$ is a dimensionless parameter (adapted from \citealt{cpp07}, see also \citealt{illarionov75}). In the simplest wind-accretion case (e.g. \citealt{edgar04}), the luminosity can be estimated as 
\begin{equation}
L_{\mathrm{X}}\approx\eta\frac{\dot{M}_{\mathrm{w}} c^2G^2M_{\mathrm{BH}}^2}{a^2 (v_{\mathrm{orb}}^2+v_{\mathrm{w}}^2)^2}
\label{eq:lum}
\end{equation}
where $\eta$ is the efficiency, $\dot{M}_{\mathrm{w}}$ is the wind mass loss rate, $a$ is the orbital separation, $v_{\mathrm{orb}}$ is the orbital velocity, and $v_{\mathrm{w}}$ is the wind velocity at the BH orbit. Assuming $\dot{M}_{\mathrm{w}} = 10^{-5}$~M$_\odot$~yr$^{-1}$ and $v_{\mathrm{w}} =1000$~km~s$^{-1}$ for the WR star (e.g. \citealt{crowther07}), $a=5.8\times 10^{11}$~cm (for a 10-$M_\odot$ companion), $M_{\mathrm{BH}}=75$~M$_\odot$, and the formation of a disc with $\eta=0.1$, the corresponding luminosity is $L_{\mathrm{X}}\simeq2\times10^{40}$~\lum. More in general, for $M_{\mathrm{BH}}>10$~M$_\odot$ and all the other things being equal, one finds $L_{\mathrm{X}}\ga3\times10^{39}$~\lum.
In case of Roche lobe overflow, even higher X-ray luminosity could be achieved. However we note that if \cgx\ is indeed a WR--BH binary, the WR star is probably not filling its Roche lobe (unless it is very massive; see for example the discussion of the case of Cyg\,X--3, where the orbital period is much shorter, in \citealt{szostek08}).
An X-ray luminosity of $\sim$$2\times10^{40}$~\lum\ can be therefore accounted for.  We finally notice that, although we do not regard the question as crucial, the problem of the lifetime of the system discussed by \citet{weisskopf04} would be significantly attenuated, since the WR phase of a massive O-type star is thought to last a few $\times$$10^5$ years \citep{meynet05}. \\

\subsection{Statistics, environment, and WR--BH binaries as ULXs}\label{demographics}

All known WR HMXBs have been mentioned in the previous Section. Three have been established so far as certain WR--BH systems:  IC\,10~X--1, NGC\,300~X--1, and M101\,ULX--1. For the fourth WR--compact object binary, Cyg\,X--3 in our Galaxy, it is still debated whether the compact object is a BH or a NS. There are however several pieces of evidence (radio, infrared and X-ray emission properties) that point to a 2--5~M$_\odot$ BH, as suggested also by evolutionary models (e.g. \citealt{lommen05,szostek08,szm08,shrader10,zdziarski13}). Apart from \cgx, discussed in this paper, two additional WR--BH binary candidates were found in the last two years: CXOU\,J123030 in NGC\,4490 and CXOU\,J004732.0--251722.1 in NGC\,253.

We expect WR--BH binaries to be associated with star forming regions, since WR are young stars with massive progenitors (with zero-age main sequence mass $\ga$25~M$_\odot$). Table~\ref{tab:summary} shows the star formation rate (SFR) of the host galaxies of WR--BH binaries and binary candidates. The average SFR is $\sim$2~M$_\odot$~yr$^{-1}$, which is quite high for nearby late-type galaxies. All metallicities listed in  Table~\ref{tab:summary} are sub-solar. Recent $N$-body and population synthesis simulations of young star clusters \citep{mapelli13,mapelli14} suggest that $\approx$2 per cent of all HMXBs powered by BHs in star forming regions are BH--WR binaries, independent of star cluster metallicity.
\begin{table*}
\begin{minipage}{14.2cm}
\centering
\caption{Properties of observed WR--BH binaries and candidates (denoted by stars), and of their host galaxies.}
\label{tab:summary}
\begin{tabular}[!h]{lccccccc}
\hline
Host galaxy & Source & Period  & BH mass$^{a}$ & WR mass$^{a}$ & SFR$^{b}$ & $Z$$^{b}$ & $t_{\rm GW}$$^{c}$ \\
 & & (h) & (M$_\odot$) & (M$_\odot$) & (M$_\odot{}$ yr$^{-1}$) &  (Z$_\odot$) & (Gyr) \\

\hline
IC\,10         & X--1                   &	34.9 & 33   & 35 & 0.07   & 0.22 & 1.4\\
NGC\,300       & X--1                   &	32.8  & 20   & 26 & 0.14   & 0.19 & 1.7 \\
NGC\,4490      & \phantom{~*}CXOU\,J123030.3+413853~* & 	6.4  &  --   & --  & 4.5    & 0.23 & 0.038\\
NGC\,253     & \phantom{~*}CXOU\,J004732.0--251722.1~* & 	14.5 &  --    & --  & 4.0    & 0.24 & 0.33\\
Circinus     & \phantom{~*}\cgx~*  &	7.2  &  --   & --  & 1.5    & 0.10 & 0.052 \\
M\,101	     & ULX--1	             & 196.8 &  20   & 19 & 3.1    & 0.17 & 200 \\
Milky~Way    & Cyg\,X--3		     & 4.8   &  3    & 7  & 0.25   & 0.31 & 0.051 \\
\hline
\end{tabular}
\begin{list}{}{}
\item[$^{a}$] For the BH and WR masses, we list only the fiducial values that we use to derive the merger rates $R$ in equation~\ref{eq:rate}. Most of these masses are very uncertain, as discussed in \cite{prestwich07,silverman08,carpano07,cpp07,crowther10,eism13,maccarone14,liu13,shrader10}. 
\item[$^{b}$] The values of SFR and metallicity of the host galaxy come from the compilation of \cite{mapelli10}. The metallicity $Z$ refers to the value at 0.7~$R_{25}$, where $R_{25}$ is the Holmberg radius of the galaxy, for NGC\,253, NGC\,300, NGC\,4490, M\,101, and for the Milky Way, while it is the total metallicity for IC\,10 and Circinus. We assume $Z_{\odot}=0.02$.
\item[$^{c}$] The parameter $t_{\rm GW}$ is the timescale for the binary to coalesce, under the assumptions discussed in the main text.
\end{list}
\end{minipage}
\end{table*}

The average luminosity of \cgx\ in the \cxo\ observations presented here makes it a bona fide ULX. The maximum luminosity reached by the source, following \citealt{weisskopf04}, is $L_{\mathrm{X}}=(1.5$--$2)\times10^{40}$~\lum. At such flux levels, in high counting statistics \xmm\ spectra, a persistent ULX typically shows the hallmark of the ultraluminous state \citep*{grd09}, with the presence of two thermal components, one of which centred at rather soft energies ($\sim$0.1--0.2~keV) and the other producing a shallow but significant rollover around 3--5~keV. Such a spectrum is usually interpreted as the imprint of a super-Eddington accretion regime (e.g. \citealt*{middleton11}).

As discussed in Section\,\ref{cgx1}, the X-ray spectrum of the \cxo\ observations considered here is satisfactorily fitted with a single component model. The lack of evidence of multiple components may be real, but may also be caused by the comparatively low statistics (the flux is $\sim$5 times lower than at maximum) and/or by the fact that \cxo\ has a lower sensitivity than \xmm\ below $\sim$0.5~keV, where the soft component peaks. If the spectrum is intrinsically single component, this may suggest that \cgx\ does not enter into the ultraluminous state and hence does not accrete above Eddington.

The latter possibility is consistent with the scenario discussed above in which accretion proceeds through a wind and the accretion rate does not exceed the Eddington limit. For wind accretion from a compact WR star, the very formation of a standard accretion disc is uncertain and the accretion efficiency can in general be smaller than that of a standard disc (e.g. \citealt*{frank02}). A scenario of rather efficient accretion from a wind of a massive WR star onto a BH of a few tens $M_\odot$ has been invoked also for M\,101~ULX--1. This source is a WR--BH ULX system with a dynamical mass measurement and a cool disc X-ray spectrum at maximum \citep{liu13}, and it has a significantly larger orbital period ($\sim$8~days, see Table\,\ref{tab:summary}) and smaller luminosity ($\sim$$ 3\times 10^{39}$\lum) than \cgx. Assuming a similar scenario also for \cgx, simple BH mass estimates based on Eddington-limited accretion from a He--WR star (as those reported above) give $M_{\mathrm{BH}}\gtrsim70$~M$_\odot$ for the observed maximum luminosity. Such a massive BH would populate the high-mass tail of the distribution of BHs formed through direct collapse of a massive star in a low metallicity environment \citep{mapelli09,zampieri09,belczynski10}, scenario indeed consistent with the metallicity inferred for the CG (see Table\,\ref{tab:summary}). In the same hypothesis, an independent limit can be obtained from the normalisation of the disc component at maximum luminosity (\cxo\ Obs. ID 365; \citealt{weisskopf04}), giving $M_{\mathrm{BH}}\simeq 8 [(d/4.2\,{\rm Mpc})/\sqrt{\cos i}]$~M$_\odot$ ($i$ is the inclination angle of the disc, e.g. \citealt{lorenzin09}).

On the other hand, the lack of a high counting statistics spectrum at maximum luminosity prevents us from reaching a robust conclusion. We thus briefly consider also the possibility that the system is accreting from a particularly massive and big WR companion via Roche-lobe overflow, for which the mass transfer rate is expected to significantly exceed the Eddington limit (e.g. \citealt{lommen05}). In this assumptions, the maximum observed luminosity of \cgx\ would place it in the populated part of the ULX luminosity distribution (e.g. \citealt{swartz11}), where the observed flux can be produced by moderately beamed, super-Eddington emission from accretion onto a BH of a few tens $M_\odot$ (or even a canonical stellar-mass BH).

It is interesting to compare the properties of the two candidate WR--BH ULX systems that we have tentatively identified, \cgx\ and CXOU\,J123030 in  NGC\,4490 \citep{eism13}. While the orbital periods are similar (7.2 and 6.4~h, respectively), the X-ray luminosity is significantly different (\cgx\ being $>10$ times more luminous). Assuming a similar scenario of sub-Eddington accretion from a wind with the formation of a disc, the different luminosity could be mostly ascribed to the different BH mass, with \cgx\ being about $\sim$10 times more massive than CXOU\,J123030. For non-extreme inclinations, this is consistent with the BH mass inferred from the normalisation of the disc component of CXOU\,J123030, for which we obtained $M_{\mathrm{BH}}\simeq 2.8 [(d/8\,{\rm Mpc})/\sqrt{\cos i}]$~M$_\odot$ \citep{eism13}. On the other hand, in case of non-standard super-Eddington accretion via Roche-lobe overflow, the larger luminosity of \cgx\ may be caused by the larger accretion rate and we would then be witnessing the different X-ray outcome produced by similar BHs in very different accretion environments.

\subsection{WR--BH binaries as precursors of BH--BH binaries}

There is great uncertainty on the expected rate of BH--BH mergers in the frequency range that will be observed by Advanced LIGO and Virgo ($\sim$10--$10^4$~Hz; \citealt{abadie10}). In fact, while for NS--NS binaries the expected merger rate can be derived from the properties of the observed NS--NS binaries (e.g. \citealt{kim03}) and from the rate of short gamma-ray bursts (e.g. \citealt{coward12,fong12}), no evidence has been found of BH--BH systems yet. 
WR--BH binaries can provide us with essential clues, since they are a possible precursor of BH--BH binaries (or BH--NS binaries), provided that the system is so tight that it remains bound when the WR star evolves into a compact remnant. 

Here, we use the properties of all known WR--BH binaries and candidates to infer the BH--BH merger rate in the instrumental range of Advanced LIGO and Virgo. The main source of uncertainty is represented by the fate of the WR: we do not know the natal kick and the mass of the compact remnant that will form from the evolution of the WR. In particular, the natal kick might unbind the binary  or transform it into a loose system that will not merge in a Hubble time. Thus, population synthesis simulations of the evolution of each WR--BH system are necessary, in order to predict the binary fate (e.g. \citealt{belczynski13}). On the other hand, our knowledge of the current orbital parameters and of the masses of the binary members is poor for most WR--BH systems: the WR star has been detected only in Cyg\,X--3, IC\,10~X--1, NGC\,300~X--1 and M101\,ULX--1, and the mass of the BH is poorly determined even in these four cases, due to the uncertainty on the inclination and because the WR winds might affect the radial velocity estimate (e.g. \citealt{maccarone14}). Given these uncertainties, population synthesis models can hardly constrain the fate of most WR--BH candidates. Thus, we adopt a much simpler approach in order to obtain an upper limit to the BH--BH merger rate from the observed WR--BH binaries and candidates. As already done by \cite{maccarone14}, we assume that all seven WR--BH candidates in our sample will become BH--BH binaries through direct collapse of the WR star (leading to a secondary BH mass $m_2=10$~M$_\odot$), and that the orbital properties of the binary will be substantially unchanged after the collapse of the WR star. When even the mass of the primary BH is not known (Table\,\ref{tab:summary}), we assume $m_1=10$~M$_\odot$.

Thus, the rate of BH--BH mergers per Mpc$^3$ ($R$) can be approximately estimated as (see e.g. \citealt{mhm10,mapelli12,ziosi14}):
\begin{equation}\label{eq:rate}
R=\rho{}_{\rm SFR}(z)\,{}\sum_i (t_{\rm GW,\,{}i}+t_{\rm evol,\,{}i})^{-1}\,{}({\rm SFR}_{i})^{-1},
\end{equation}
where $\rho{}_{\rm SFR}(z)$ is the cosmic SFR density \citep{hopkins06}, $t_{\rm GW,\,{}i}$ is the coalescence timescale of the $i-$th binary ($i=1,..,7$ in our sample, see Table~\ref{tab:summary}), $t_{\rm evol,\,{}i}$ is the time elapsed from the formation of the $i-$th binary (as a binary of two main sequence stars) to the birth of the second BH, and SFR$_i$ is the current SFR of the $i-$th galaxy. For our sample, we assume $t_{\rm evol\,{}i}\approx{}3\times{}10^6$~yr, which is the main sequence lifetime of the most massive stars. The values of SFR$_i$ are from Table~\ref{tab:summary}. Equation~\ref{eq:rate} gives a strong upper limit for the sample, for the fact that we neglect local galaxies that do not host any WR--BH binary.

The coalescence timescale $t_{\rm GW}$ can be expressed as \citep{peters64}:
\begin{equation}
t_{\rm GW}=\frac{5}{256}\frac{c^5\,{}a^4\,{}(1-e^2)^{7/2}}{G^3\,{}m_1\,{}m_2\,{}(m_1+m_2)},
\end{equation}
where $c$ is the light speed, $G$ is the gravitational constant, $a$ is the semi-major axis and $e$ the eccentricity of the orbit. We assume  $e=0$ for all considered binaries. Table~\ref{tab:summary} shows the values of $t_{\rm GW}$ for each considered binary, under these assumptions. The value of $t_{\rm GW}$ for M\,101~ULX--1 is much larger than the Hubble time: the impact of this system is negligible for any estimate of the Advanced LIGO and Virgo detection rate. The second longest value of $t_{\rm GW}$ is $\sim$2~Gyr (for NGC\,300~X--1): this value is well below the Hubble time, and implies that we might detect at present time the merger of systems like NGC\,300~X--1 that formed $\sim$2~Gyr ago, i.e. at redshift $z=0.3$, when the SFR density was a factor of two higher than now [$\rho{}_{\rm SFR}(z=0.3)\sim{}3.6\times{}10^{-2}$ M$_\odot$ yr$^{-1}$ Mpc$^{-3}$, \citealt{hopkins06}].  To obtain the best favourable upper limit, we use $\rho{}_{\rm SFR}(z=0.3)=3.6\times{}10^{-2}$ M$_\odot$ yr$^{-1}$ Mpc$^{-3}$ for all sources in equation~\ref{eq:rate}, and we obtain $R\sim{}4\times{}10^{-9}$ and $1\times{}10^{-9}$ yr$^{-1}$ Mpc$^{-3}$, if we include or do not include Cyg\,X--3 among the WR--BH candidates, respectively. Since Advanced LIGO and Virgo will detect BH--BH mergers out to $\sim{}1$ Gpc, we can infer upper limits to the detection rate of BH--BH mergers of $\sim$16 and 5 events per year,  if we include or do not include Cyg\,X--3, respectively. Such rates are consistent with those predicted in previous papers that study WR--BH binaries (e.g. \citealt{belczynski13,maccarone14}). It is worth noting that our upper limit is quite close to the `pessimistic' estimate of the BH--BH merger rate by the Advanced LIGO--Virgo collaboration (whose pessimistic and optimistic rates are $R=10^{-10}$ and $3\times{}10^{-7}$~yr$^{-1}$~Mpc$^{-3}$, respectively; \citealt{abadie10}).

\section{Summary and conclusions}\label{summary}

We reported on our timing survey of the deep \cxo/ACIS observations of the Circinus galaxy and its surroundings (Fig.\,\ref{circinus_ds9}). Approximately 150 X-ray sources were detected and for about 40 of them, enough photons were collected to search for periodic signals by means of a Fourier transform. We discovered two new X-ray pulsators, \ip\ and the uncatalogued \cv, and bumped into the already known periodic modulation of the notable and controversial source \cgx\  ($P\simeq7.2$~h; \citealt{bauer01}). For the other sources, we set upper limits on the presence of periodic signals (Fig.\,\ref{ulimits}). 

\ip\ is more than 2~arcmin out of the border of the CG. A spin period of 1.7~h and an orbital period of 17.8~h give the source away as an intermediate polar seen at large inclination. Its X-ray spectrum can be modelled by a power-law with photon index $\Gamma\simeq1.4$. The flux in the longest \cxo\ observation was $\approx$$1\times10^{-13}$~\flux\ and other observations imply a variability of $\approx$50~per~cent on time-scales of weeks--years. The typical luminosity of IPs and the nondetection of the optical counterpart suggest a distance larger than $\sim$5~kpc. 

Albeit at $\sim$3.5~arcmin from the nucleus, \cv\ appears inside the CG. However, the low absorption column, which is much smaller than the total Galactic density, argues against an extragalactic source. Indeed, the probability of a foreground Galactic X-ray source is  substantial ($\approx$10~per~cent). The period of this source is 1.8~h and the spectrum is hard: it can be described by a power law with $\Gamma\simeq0.9$. The observed flux was $\approx$$5\times10^{-14}$~\flux, with $\approx$50~per~cent variations on weekly/yearly scales.  We believe that also \cv\ is a Galactic magnetic cataclysmic variable, most probably of the polar type. Assuming that the companion is a M5V star (or similar), the nondetection of its optical counterpart is not particularly constraining for the distance, implying only $d \ga 0.7$~kpc. On the other hand, if the system is within a few kpc, its luminosity is in the normal range for polars.

We deem the association of \cgx\ with the CG convincing. Prompted by the similarity of its modulation to the distinctive ones of the known WR--HMXRBs and candidates, we advanced the possibility that \cgx\ might be one of such systems. The observations of \cgx\ (in particular, the high luminosity implied and the limits on the optical counterpart) are consistent with this hypothesis. The dearth of observed WR--HMXRBs is puzzling, since they should be relatively common and very bright (\citealt{lommen05}; \citealt*{linden12}). Though there is no obvious bias against their detection, it is possible that some of these objects might be misclassified or unrecognised. At any rate, in the last few years, the sample of WR--HMXRBs has grown rapidly to four confirmed sources and three candidates (including \cgx), all of them outside our Galaxy, with the exception of Cyg\,X--3. Besides their relevance for the population of X-ray binaries and ULXs, WR--BH systems are very important as they might be progenitors of double-BH systems. We used the information from the current sample of systems to estimate an upper limit to the detection rate of stellar BH--BH mergers with Advanced LIGO and Virgo, which turned out to be $\sim$16~yr$^{-1}$ for a distance range of 1~Gpc.

\section*{Acknowledgements} 
This research is based on data obtained from the \cxo\ Data Archive and has made use of software provided by the \cxo\ X-ray Center (CXC) in the application package \textsc{ciao}. This research has also made use of data obtained from the ESA's \xmm\ Science Archive (XSA) and from the ESO Science Archive Facility (under request number 161735), and of the NED, which is operated by the JPL, Caltech, under contract with the NASA.
The \textsc{iraf} software is distributed by the NOAO, which is operated by AURA, Inc., under cooperative agreement with the NSF.
PE acknowledges a Fulbright Research Scholar grant administered by the U.S.--Italy Fulbright Commission and is grateful to the Harvard--Smithsonian Center for Astrophysics for hosting him during his Fulbright exchange. 
MM acknowledges financial support from the MIUR through grant FIRB 2012 RBFR12PM1F, and from INAF through grants PRIN-2011-1 and PRIN-2014-14.
LZ acknowledges financial support from the ASI/INAF contract n. I/037/12/0. LS acknowledges the PRIN-INAF 2014 grant `Towards a unified picture of accretion in High Mass X-Ray Binaries'.
PE thanks M. Mezcua for surveying the available \hst\ data of the Circinus region and A. Wolter for comments on the manuscript.

\bibliographystyle{mn2e}
\bibliography{biblio}

\begin{thebibliography}{120}
\expandafter\ifx\csname natexlab\endcsname\relax\def\natexlab#1{#1}\fi

\bibitem[{{Abadie} {et~al}\mbox{.}(2010){Abadie}, {Abbott}, {Abbott},
  {Abernathy}, {Accadia}, {Acernese}, {Adams}, {Adhikari}, {Ajith}, {Allen}, \&
  et~al.}]{abadie10}
{Abadie} J. {et~al.}, 2010, Classical and Quantum Gravity, 27, 173001

\bibitem[{{Aizu}(1973)}]{aizu73}
{Aizu} K., 1973, Progress of Theoretical Physics, 49, 1184

\bibitem[{{Anders} \& {Grevesse}(1989)}]{anders89}
{Anders} E., {Grevesse} N., 1989, \gca, 53, 197

\bibitem[{{Ar{\'e}valo} {et~al}\mbox{.}(2014){Ar{\'e}valo}, {Bauer},
  {Puccetti}, {Walton}, {Koss}, {Boggs}, {Brandt}, {Brightman}, {Christensen},
  {Comastri}, {Craig}, {Fuerst}, {Gandhi}, {Grefenstette}, {Hailey},
  {Harrison}, {Luo}, {Madejski}, {Madsen}, {Marinucci}, {Matt}, {Saez},
  {Stern}, {Stuhlinger}, {Treister}, {Urry}, \& {Zhang}}]{arevalo14}
{Ar{\'e}valo} P. {et~al.}, 2014, \apj, 791, 81

\bibitem[{{Bachetti} {et~al}\mbox{.}(2014){Bachetti}, {Harrison}, {Walton},
  {Grefenstette}, {Chakrabarty}, {F{\"u}rst}, {Barret}, {Beloborodov}, {Boggs},
  {Christensen}, {Craig}, {Fabian}, {Hailey}, {Hornschemeier}, {Kaspi},
  {Kulkarni}, {Maccarone}, {Miller}, {Rana}, {Stern}, {Tendulkar}, {Tomsick},
  {Webb}, \& {Zhang}}]{bachetti14}
{Bachetti} M. {et~al.}, 2014, \nat, 514, 202

\bibitem[{{Balucinska-Church} \& {McCammon}(1992)}]{balucinska92}
{Balucinska-Church} M., {McCammon} D., 1992, \apj, 400, 699

\bibitem[{{Bauer} {et~al}\mbox{.}(2001){Bauer}, {Brandt}, {Sambruna},
  {Chartas}, {Garmire}, {Kaspi}, \& {Netzer}}]{bauer01}
{Bauer} F.~E., {Brandt} W.~N., {Sambruna} R.~M., {Chartas} G., {Garmire} G.~P.,
  {Kaspi} S., {Netzer} H., 2001, \aj, 122, 182

\bibitem[{{Bauer} {et~al}\mbox{.}(2008){Bauer}, {Dwarkadas}, {Brandt},
  {Immler}, {Smartt}, {Bartel}, \& {Bietenholz}}]{bauer08}
{Bauer} F.~E., {Dwarkadas} V.~V., {Brandt} W.~N., {Immler} S., {Smartt} S.,
  {Bartel} N., {Bietenholz} M.~F., 2008, \apj, 688, 1210

\bibitem[{{Belczynski} {et~al}\mbox{.}(2010){Belczynski}, {Bulik}, {Fryer},
  {Ruiter}, {Valsecchi}, {Vink}, \& {Hurley}}]{belczynski10}
{Belczynski} K., {Bulik} T., {Fryer} C.~L., {Ruiter} A., {Valsecchi} F., {Vink}
  J.~S., {Hurley} J.~R., 2010, \apj, 714, 1217

\bibitem[{{Belczynski} {et~al}\mbox{.}(2013){Belczynski}, {Bulik}, {Mandel},
  {Sathyaprakash}, {Zdziarski}, \& {Miko{\l}ajewska}}]{belczynski13}
{Belczynski} K., {Bulik} T., {Mandel} I., {Sathyaprakash} B.~S., {Zdziarski}
  A.~A., {Miko{\l}ajewska} J., 2013, \apj, 764, 96

\bibitem[{{Berghea} {et~al}\mbox{.}(2008){Berghea}, {Weaver}, {Colbert}, \&
  {Roberts}}]{berghea08}
{Berghea} C.~T., {Weaver} K.~A., {Colbert} E.~J.~M., {Roberts} T.~P., 2008,
  \apj, 687, 471

\bibitem[{{Bianchi} {et~al}\mbox{.}(2002){Bianchi}, {Matt}, {Fiore}, {Fabian},
  {Iwasawa}, \& {Nicastro}}]{bianchi02}
{Bianchi} S., {Matt} G., {Fiore} F., {Fabian} A.~C., {Iwasawa} K., {Nicastro}
  F., 2002, \aap, 396, 793

\bibitem[{{Bilir} {et~al}\mbox{.}(2008){Bilir}, {Karaali}, {Ak}, {Yaz},
  {Cabrera-Lavers}, \& {Co{\c s}kuno{\v g}lu}}]{bilir08}
{Bilir} S., {Karaali} S., {Ak} S., {Yaz} E., {Cabrera-Lavers} A., {Co{\c
  s}kuno{\v g}lu} K.~B., 2008, \mnras, 390, 1569

\bibitem[{{Bochanski} {et~al}\mbox{.}(2011){Bochanski}, {Hawley}, \&
  {West}}]{bochanski11}
{Bochanski} J.~J., {Hawley} S.~L., {West} A.~A., 2011, \aj, 141, 98

\bibitem[{{Carpano} {et~al}\mbox{.}(2007{\natexlab{a}}){Carpano}, {Pollock},
  {Prestwich}, {Crowther}, {Wilms}, {Yungelson}, \& {Ehle}}]{cpp07}
{Carpano} S., {Pollock} A.~M.~T., {Prestwich} A., {Crowther} P., {Wilms} J.,
  {Yungelson} L., {Ehle} M., 2007{\natexlab{a}}, \aap, 466, L17

\bibitem[{{Carpano} {et~al}\mbox{.}(2007{\natexlab{b}}){Carpano}, {Pollock},
  {Wilms}, {Ehle}, \& {Schirmer}}]{carpano07}
{Carpano} S., {Pollock} A.~M.~T., {Wilms} J., {Ehle} M., {Schirmer} M.,
  2007{\natexlab{b}}, \aap, 461, L9

\bibitem[{{Coward} {et~al}\mbox{.}(2012){Coward}, {Howell}, {Piran}, {Stratta},
  {Branchesi}, {Bromberg}, {Gendre}, {Burman}, \& {Guetta}}]{coward12}
{Coward} D.~M. {et~al.}, 2012, \mnras, 425, 2668

\bibitem[{{Crowther}(2007)}]{crowther07}
{Crowther} P.~A., 2007, \araa, 45, 177

\bibitem[{{Crowther} {et~al}\mbox{.}(2010){Crowther}, {Barnard}, {Carpano},
  {Clark}, {Dhillon}, \& {Pollock}}]{crowther10}
{Crowther} P.~A., {Barnard} R., {Carpano} S., {Clark} J.~S., {Dhillon} V.~S.,
  {Pollock} A.~M.~T., 2010, \mnras, 403, L41

\bibitem[{{Davis}(2001)}]{davis01}
{Davis} J.~E., 2001, \apj, 562, 575

\bibitem[{{De Luca} \& {Molendi}(2004)}]{deluca04}
{De Luca} A., {Molendi} S., 2004, \aap, 419, 837

\bibitem[{{D{\'{\i}}az Trigo} {et~al}\mbox{.}(2006){D{\'{\i}}az Trigo},
  {Parmar}, {Boirin}, {M{\'e}ndez}, \& {Kaastra}}]{diaztrigo06}
{D{\'{\i}}az Trigo} M., {Parmar} A.~N., {Boirin} L., {M{\'e}ndez} M., {Kaastra}
  J.~S., 2006, \aap, 445, 179

\bibitem[{{Dickey} \& {Lockman}(1990)}]{dickey90}
{Dickey} J.~M., {Lockman} F.~J., 1990, \araa, 28, 215

\bibitem[{{Drew} {et~al}\mbox{.}(2014){Drew}, {Gonzalez-Solares}, {Greimel},
  {Irwin}, {K{\"u}pc{\"u} Yoldas}, {Lewis}, {Barentsen}, {Eisl{\"o}ffel},
  {Farnhill}, {Martin}, {Walsh}, {Walton}, {Mohr-Smith}, {Raddi}, {Sale},
  {Wright}, {Groot}, {Barlow}, {Corradi}, {Drake}, {Fabregat}, {Frew},
  {G{\"a}nsicke}, {Knigge}, {Mampaso}, {Morris}, {Naylor}, {Parker},
  {Phillipps}, {Ruhland}, {Steeghs}, {Unruh}, {Vink}, {Wesson}, \&
  {Zijlstra}}]{drew14}
{Drew} J.~E. {et~al.}, 2014, \mnras, 440, 2036

\bibitem[{{Edgar}(2004)}]{edgar04}
{Edgar} R., 2004, \nar, 48, 843

\bibitem[{{Esposito} {et~al}\mbox{.}(2015){Esposito}, {Israel}, {de Martino},
  {D'Avanzo}, {Testa}, {Sidoli}, {Di Stefano}, {Belfiore}, {Mapelli},
  {Piranomonte}, {Rodr{\'{\i}}guez Castillo}, {Moretti}, {D'Elia},
  {Verrecchia}, {Campana}, \& {Rea}}]{esposito15}
{Esposito} P. {et~al.}, 2015, \mnras, 450, 1705

\bibitem[{{Esposito} {et~al}\mbox{.}(2013{\natexlab{a}}){Esposito}, {Israel},
  {Sidoli}, {Mapelli}, {Zampieri}, \& {Motta}}]{eism13}
{Esposito} P., {Israel} G.~L., {Sidoli} L., {Mapelli} M., {Zampieri} L.,
  {Motta} S.~E., 2013{\natexlab{a}}, \mnras, 436, 3380

\bibitem[{{Esposito} {et~al}\mbox{.}(2013{\natexlab{b}}){Esposito}, {Israel},
  {Sidoli}, {Mason}, {Rodr{\'{\i}}guez Castillo}, {Halpern}, {Moretti}, \&
  {G{\"o}tz}}]{eis13}
{Esposito} P., {Israel} G.~L., {Sidoli} L., {Mason} E., {Rodr{\'{\i}}guez
  Castillo} G.~A., {Halpern} J.~P., {Moretti} A., {G{\"o}tz} D.,
  2013{\natexlab{b}}, \mnras, 433, 2028

\bibitem[{{Esposito} {et~al}\mbox{.}(2013{\natexlab{c}}){Esposito}, {Israel},
  {Sidoli}, {Rodr{\'{\i}}guez Castillo}, {Masetti}, {D'Avanzo}, \&
  {Campana}}]{eisrc13}
{Esposito} P., {Israel} G.~L., {Sidoli} L., {Rodr{\'{\i}}guez Castillo} G.~A.,
  {Masetti} N., {D'Avanzo} P., {Campana} S., 2013{\natexlab{c}}, \mnras, 433,
  3464

\bibitem[{{Esposito} {et~al}\mbox{.}(2014){Esposito}, {Israel}, {Sidoli},
  {Tiengo}, {Campana}, \& {Moretti}}]{eis14}
{Esposito} P., {Israel} G.~L., {Sidoli} L., {Tiengo} A., {Campana} S.,
  {Moretti} A., 2014, \mnras, 441, 1126

\bibitem[{{Esposito} {et~al}\mbox{.}(2011){Esposito}, {Turolla}, {De Luca},
  {Israel}, {Possenti}, \& {Burrows}}]{etdl11}
{Esposito} P., {Turolla} R., {De Luca} A., {Israel} G.~L., {Possenti} A.,
  {Burrows} D.~N., 2011, \mnras, 418, 170

\bibitem[{{Fabbiano}(2006)}]{fabbiano06}
{Fabbiano} G., 2006, \araa, 44, 323

\bibitem[{{Feng} \& {Soria}(2011)}]{feng11}
{Feng} H., {Soria} R., 2011, \nar, 55, 166

\bibitem[{{Fong} {et~al}\mbox{.}(2012){Fong}, {Berger}, {Margutti}, {Zauderer},
  {Troja}, {Czekala}, {Chornock}, {Gehrels}, {Sakamoto}, {Fox}, \&
  {Podsiadlowski}}]{fong12}
{Fong} W. {et~al.}, 2012, \apj, 756, 189

\bibitem[{{Fragos} {et~al}\mbox{.}(2015){Fragos}, {Linden}, {Kalogera}, \&
  {Sklias}}]{fragos15}
{Fragos} T., {Linden} T., {Kalogera} V., {Sklias} P., 2015, \apjl, 802, L5

\bibitem[{{Frank} {et~al}\mbox{.}(2002){Frank}, {King}, \& {Raine}}]{frank02}
{Frank} J., {King} A., {Raine} D.~J., 2002, {Accretion Power in Astrophysics:
  Third Edition}. Cambridge: Cambridge University Press

\bibitem[{{Freeman} {et~al}\mbox{.}(1977){Freeman}, {Karlsson}, {Lynga},
  {Burrell}, {van Woerden}, {Goss}, \& {Mebold}}]{freeman77}
{Freeman} K.~C., {Karlsson} B., {Lynga} G., {Burrell} J.~F., {van Woerden} H.,
  {Goss} W.~M., {Mebold} U., 1977, \aap, 55, 445

\bibitem[{{Fruscione} {et~al}\mbox{.}(2006){Fruscione}, {McDowell}, {Allen},
  {Brickhouse}, {Burke}, {Davis}, {Durham}, {Elvis}, {Galle}, {Harris},
  {Huenemoerder}, {Houck}, {Ishibashi}, {Karovska}, {Nicastro}, {Noble},
  {Nowak}, {Primini}, {Siemiginowska}, {Smith}, \& {Wise}}]{fruscione06}
{Fruscione} A. {et~al.}, 2006, in SPIE Conference Series, Vol. 6270,
  Observatory Operations: Strategies, Processes, and Systems, {Silva} D.~R.,
  {Doxsey} R.~E., eds., SPIE, Bellingham, p. 62701V

\bibitem[{{Garmire} {et~al}\mbox{.}(2003){Garmire}, {Bautz}, {Ford}, {Nousek},
  \& {Ricker}}]{garmire03}
{Garmire} G.~P., {Bautz} M.~W., {Ford} P.~G., {Nousek} J.~A., {Ricker}, Jr.
  G.~R., 2003, in Proceedings of the SPIE., Vol. 4851, X-Ray and Gamma-Ray
  Telescopes and Instruments for Astronomy., {Truemper} J.~E., {Tananbaum}
  H.~D., eds., SPIE, Bellingham, pp. 28--44

\bibitem[{{Gladstone} {et~al}\mbox{.}(2013){Gladstone}, {Copperwheat},
  {Heinke}, {Roberts}, {Cartwright}, {Levan}, \& {Goad}}]{gladstone13}
{Gladstone} J.~C., {Copperwheat} C., {Heinke} C.~O., {Roberts} T.~P.,
  {Cartwright} T.~F., {Levan} A.~J., {Goad} M.~R., 2013, \apjs, 206, 14

\bibitem[{{Gladstone} {et~al}\mbox{.}(2009){Gladstone}, {Roberts}, \&
  {Done}}]{grd09}
{Gladstone} J.~C., {Roberts} T.~P., {Done} C., 2009, \mnras, 397, 1836

\bibitem[{{Guainazzi} {et~al}\mbox{.}(1999){Guainazzi}, {Matt}, {Antonelli},
  {Bassani}, {Fabian}, {Maiolino}, {Marconi}, {Fiore}, {Iwasawa}, \&
  {Piro}}]{guainazzi99}
{Guainazzi} M. {et~al.}, 1999, \mnras, 310, 10

\bibitem[{{Haberl}(1991)}]{haberl91}
{Haberl} F., 1991, \aap, 252, 272

\bibitem[{{Hertz} {et~al}\mbox{.}(1978){Hertz}, {Joss}, \&
  {Rappaport}}]{hertz78}
{Hertz} P., {Joss} P.~C., {Rappaport} S., 1978, \apj, 224, 614

\bibitem[{{Hopkins} \& {Beacom}(2006)}]{hopkins06}
{Hopkins} A.~M., {Beacom} J.~F., 2006, \apj, 651, 142

\bibitem[{{Illarionov} \& {Sunyaev}(1975)}]{illarionov75}
{Illarionov} A.~F., {Sunyaev} R.~A., 1975, \aap, 39, 185

\bibitem[{{Israel} \& {Stella}(1996)}]{israel96}
{Israel} G.~L., {Stella} L., 1996, \apj, 468, 369

\bibitem[{{Kalberla} {et~al}\mbox{.}(2005){Kalberla}, {Burton}, {Hartmann},
  {Arnal}, {Bajaja}, {Morras}, \& {P{\"o}ppel}}]{kalberla05}
{Kalberla} P.~M.~W., {Burton} W.~B., {Hartmann} D., {Arnal} E.~M., {Bajaja} E.,
  {Morras} R., {P{\"o}ppel} W.~G.~L., 2005, \aap, 440, 775

\bibitem[{{Kim} {et~al}\mbox{.}(2003){Kim}, {Kalogera}, \& {Lorimer}}]{kim03}
{Kim} C., {Kalogera} V., {Lorimer} D.~R., 2003, \apj, 584, 985

\bibitem[{{King}(2009)}]{king09}
{King} A.~R., 2009, \mnras, 393, L41

\bibitem[{{Kuijken}(2011)}]{kuijken11}
{Kuijken} K., 2011, The Messenger, 146, 8

\bibitem[{{Langer}(1989)}]{langer89}
{Langer} N., 1989, \aap, 210, 93

\bibitem[{{Leahy} {et~al}\mbox{.}(1983){Leahy}, {Darbro}, {Elsner},
  {Weisskopf}, {Kahn}, {Sutherland}, \& {Grindlay}}]{leahy83}
{Leahy} D.~A., {Darbro} W., {Elsner} R.~F., {Weisskopf} M.~C., {Kahn} S.,
  {Sutherland} P.~G., {Grindlay} J.~E., 1983, \apj, 266, 160

\bibitem[{{Linden} {et~al}\mbox{.}(2012){Linden}, {Valsecchi}, \&
  {Kalogera}}]{linden12}
{Linden} T., {Valsecchi} F., {Kalogera} V., 2012, \apj, 748, 114

\bibitem[{{Liu} {et~al}\mbox{.}(2013){Liu}, {Bregman}, {Bai}, {Justham}, \&
  {Crowther}}]{liu13}
{Liu} J.-F., {Bregman} J.~N., {Bai} Y., {Justham} S., {Crowther} P., 2013,
  \nat, 503, 500

\bibitem[{{Liu} \& {Mirabel}(2005)}]{lm05}
{Liu} Q.~Z., {Mirabel} I.~F., 2005, \aap, 429, 1125

\bibitem[{{Lommen} {et~al}\mbox{.}(2005){Lommen}, {Yungelson}, {van den
  Heuvel}, {Nelemans}, \& {Portegies Zwart}}]{lommen05}
{Lommen} D., {Yungelson} L., {van den Heuvel} E., {Nelemans} G., {Portegies
  Zwart} S., 2005, \aap, 443, 231

\bibitem[{{Lorenzin} \& {Zampieri}(2009)}]{lorenzin09}
{Lorenzin} A., {Zampieri} L., 2009, \mnras, 394, 1588

\bibitem[{{Maccarone} {et~al}\mbox{.}(2014){Maccarone}, {Lehmer}, {Leyder},
  {Antoniou}, {Hornschemeier}, {Ptak}, {Wik}, \& {Zezas}}]{maccarone14}
{Maccarone} T.~J., {Lehmer} B.~D., {Leyder} J.~C., {Antoniou} V.,
  {Hornschemeier} A., {Ptak} A., {Wik} D., {Zezas} A., 2014, \mnras, 439, 3064

\bibitem[{{Makishima} {et~al}\mbox{.}(2000){Makishima}, {Kubota}, {Mizuno},
  {Ohnishi}, {Tashiro}, {Aruga}, {Asai}, {Dotani}, {Mitsuda}, {Ueda}, {Uno},
  {Yamaoka}, {Ebisawa}, {Kohmura}, \& {Okada}}]{makishima00}
{Makishima} K. {et~al.}, 2000, \apj, 535, 632

\bibitem[{{Mapelli} {et~al}\mbox{.}(2009){Mapelli}, {Colpi}, \&
  {Zampieri}}]{mapelli09}
{Mapelli} M., {Colpi} M., {Zampieri} L., 2009, \mnras, 395, L71

\bibitem[{{Mapelli} {et~al}\mbox{.}(2010{\natexlab{a}}){Mapelli}, {Huwyler},
  {Mayer}, {Jetzer}, \& {Vecchio}}]{mhm10}
{Mapelli} M., {Huwyler} C., {Mayer} L., {Jetzer} P., {Vecchio} A.,
  2010{\natexlab{a}}, \apj, 719, 987

\bibitem[{{Mapelli} {et~al}\mbox{.}(2012){Mapelli}, {Ripamonti}, {Vecchio},
  {Graham}, \& {Gualandris}}]{mapelli12}
{Mapelli} M., {Ripamonti} E., {Vecchio} A., {Graham} A.~W., {Gualandris} A.,
  2012, \aap, 542, A102

\bibitem[{{Mapelli} {et~al}\mbox{.}(2010{\natexlab{b}}){Mapelli}, {Ripamonti},
  {Zampieri}, {Colpi}, \& {Bressan}}]{mapelli10}
{Mapelli} M., {Ripamonti} E., {Zampieri} L., {Colpi} M., {Bressan} A.,
  2010{\natexlab{b}}, \mnras, 408, 234

\bibitem[{{Mapelli} \& {Zampieri}(2014)}]{mapelli14}
{Mapelli} M., {Zampieri} L., 2014, \apj, 794, 7

\bibitem[{{Mapelli} {et~al}\mbox{.}(2013){Mapelli}, {Zampieri}, {Ripamonti}, \&
  {Bressan}}]{mapelli13}
{Mapelli} M., {Zampieri} L., {Ripamonti} E., {Bressan} A., 2013, \mnras, 429,
  2298

\bibitem[{{Mason} \& {Cordova}(1982)}]{mason82}
{Mason} K.~O., {Cordova} F.~A., 1982, \apj, 262, 253

\bibitem[{{Massaro} {et~al}\mbox{.}(2006){Massaro}, {Bianchi}, {Matt},
  {D'Onofrio}, \& {Nicastro}}]{massaro06}
{Massaro} F., {Bianchi} S., {Matt} G., {D'Onofrio} E., {Nicastro} F., 2006,
  \aap, 455, 153

\bibitem[{{Massey}(2003)}]{massey03}
{Massey} P., 2003, \araa, 41, 15

\bibitem[{{Matt} {et~al}\mbox{.}(1996){Matt}, {Fiore}, {Perola}, {Piro},
  {Fink}, {Grandi}, {Matsuoka}, {Oliva}, \& {Salvati}}]{matt96}
{Matt} G. {et~al.}, 1996, \mnras, 281, L69

\bibitem[{{Mattana} {et~al}\mbox{.}(2006){Mattana}, {G{\"o}tz}, {Falanga},
  {Senziani}, {de Luca}, {Esposito}, \& {Caraveo}}]{mattana06}
{Mattana} F., {G{\"o}tz} D., {Falanga} M., {Senziani} F., {de Luca} A.,
  {Esposito} P., {Caraveo} P.~A., 2006, \aap, 460, L1

\bibitem[{{Meynet} \& {Maeder}(2005)}]{meynet05}
{Meynet} G., {Maeder} A., 2005, \aap, 429, 581

\bibitem[{{Middleton} {et~al}\mbox{.}(2011){Middleton}, {Sutton}, \&
  {Roberts}}]{middleton11}
{Middleton} M.~J., {Sutton} A.~D., {Roberts} T.~P., 2011, \mnras, 417, 464

\bibitem[{{Mingo} {et~al}\mbox{.}(2012){Mingo}, {Hardcastle}, {Croston},
  {Evans}, {Kharb}, {Kraft}, \& {Lenc}}]{mingo12}
{Mingo} B., {Hardcastle} M.~J., {Croston} J.~H., {Evans} D.~A., {Kharb} P.,
  {Kraft} R.~P., {Lenc} E., 2012, \apj, 758, 95

\bibitem[{{Mitsuda} {et~al}\mbox{.}(1984){Mitsuda}, {Inoue}, {Koyama},
  {Makishima}, {Matsuoka}, {Ogawara}, {Suzuki}, {Tanaka}, {Shibazaki}, \&
  {Hirano}}]{mitsuda84}
{Mitsuda} K. {et~al.}, 1984, \pasj, 36, 741

\bibitem[{{Molendi} {et~al}\mbox{.}(2003){Molendi}, {Bianchi}, \&
  {Matt}}]{molendi03}
{Molendi} S., {Bianchi} S., {Matt} G., 2003, \mnras, 343, L1

\bibitem[{{Mushotzky}(2004)}]{mushotzky04}
{Mushotzky} R., 2004, Progress of Theoretical Physics Supplement, 155, 27

\bibitem[{{Okazaki} \& {Russell}(2014)}]{okazaki14}
{Okazaki} A.~T., {Russell} C.~M.~P., 2014, in Suzaku-MAXI 2014: Expanding the
  Frontiers of the X-ray Universe, {Ishida} M., {Petre} R., {Mitsuda} K., eds.,
  p. 202

\bibitem[{{Parker} {et~al}\mbox{.}(2005){Parker}, {Norton}, \&
  {Mukai}}]{parker05}
{Parker} T.~L., {Norton} A.~J., {Mukai} K., 2005, \aap, 439, 213

\bibitem[{{Patterson}(1994)}]{patterson94}
{Patterson} J., 1994, \pasp, 106, 209

\bibitem[{{Peters}(1964)}]{peters64}
{Peters} P.~C., 1964, Physical Review, 136, 1224

\bibitem[{{Predehl} \& {Schmitt}(1995)}]{predehl95}
{Predehl} P., {Schmitt} J.~H.~M.~M., 1995, \aap, 293, 889

\bibitem[{{Prestwich} {et~al}\mbox{.}(2007){Prestwich}, {Kilgard}, {Crowther},
  {Carpano}, {Pollock}, {Zezas}, {Saar}, {Roberts}, \& {Ward}}]{prestwich07}
{Prestwich} A.~H. {et~al.}, 2007, \apjl, 669, L21

\bibitem[{{Ptak} {et~al}\mbox{.}(2006){Ptak}, {Colbert}, {van der Marel},
  {Roye}, {Heckman}, \& {Towne}}]{ptak06}
{Ptak} A., {Colbert} E., {van der Marel} R.~P., {Roye} E., {Heckman} T.,
  {Towne} B., 2006, \apjs, 166, 154

\bibitem[{{Remillard} \& {McClintock}(2006)}]{remillard06}
{Remillard} R.~A., {McClintock} J.~E., 2006, \araa, 44, 49

\bibitem[{{Ritter} \& {Kolb}(2003)}]{ritter03}
{Ritter} H., {Kolb} U., 2003, \aap, 404, 301

\bibitem[{{Sambruna} {et~al}\mbox{.}(2001){Sambruna}, {Brandt}, {Chartas},
  {Netzer}, {Kaspi}, {Garmire}, {Nousek}, \& {Weaver}}]{sambruna01}
{Sambruna} R.~M., {Brandt} W.~N., {Chartas} G., {Netzer} H., {Kaspi} S.,
  {Garmire} G.~P., {Nousek} J.~A., {Weaver} K.~A., 2001, \apjl, 546, L9

\bibitem[{{Sazonov} {et~al}\mbox{.}(2006){Sazonov}, {Revnivtsev}, {Gilfanov},
  {Churazov}, \& {Sunyaev}}]{sazonov06}
{Sazonov} S., {Revnivtsev} M., {Gilfanov} M., {Churazov} E., {Sunyaev} R.,
  2006, \aap, 450, 117

\bibitem[{{Schaerer} \& {Maeder}(1992)}]{schraerer92}
{Schaerer} D., {Maeder} A., 1992, \aap, 263, 129

\bibitem[{{Shrader} {et~al}\mbox{.}(2010){Shrader}, {Titarchuk}, \&
  {Shaposhnikov}}]{shrader10}
{Shrader} C.~R., {Titarchuk} L., {Shaposhnikov} N., 2010, \apj, 718, 488

\bibitem[{{Shu} {et~al}\mbox{.}(2011){Shu}, {Yaqoob}, \& {Wang}}]{shu11}
{Shu} X.~W., {Yaqoob} T., {Wang} J.~X., 2011, \apj, 738, 147

\bibitem[{{Silverman} \& {Filippenko}(2008)}]{silverman08}
{Silverman} J.~M., {Filippenko} A.~V., 2008, \apjl, 678, L17

\bibitem[{{Skrutskie} {et~al}\mbox{.}(2006){Skrutskie}, {Cutri}, {Stiening},
  {Weinberg}, {Schneider}, {Carpenter}, {Beichman}, {Capps}, {Chester},
  {Elias}, {Huchra}, {Liebert}, {Lonsdale}, {Monet}, {Price}, {Seitzer},
  {Jarrett}, {Kirkpatrick}, {Gizis}, {Howard}, {Evans}, {Fowler}, {Fullmer},
  {Hurt}, {Light}, {Kopan}, {Marsh}, {McCallon}, {Tam}, {Van Dyk}, \&
  {Wheelock}}]{skrutskie06}
{Skrutskie} M.~F. {et~al.}, 2006, \aj, 131, 1163

\bibitem[{{Smith} \& {Dhillon}(1998)}]{smith98}
{Smith} D.~A., {Dhillon} V.~S., 1998, \mnras, 301, 767

\bibitem[{{Smith} \& {Wilson}(2001)}]{sw01}
{Smith} D.~A., {Wilson} A.~S., 2001, \apj, 557, 180

\bibitem[{{Smith}(2006)}]{smith06}
{Smith} R.~C., 2006, Contemporary Physics, 47, 363

\bibitem[{{Somero} {et~al}\mbox{.}(2012){Somero}, {Hakala}, {Muhli}, {Charles},
  \& {Vilhu}}]{somero12}
{Somero} A., {Hakala} P., {Muhli} P., {Charles} P., {Vilhu} O., 2012, \aap,
  539, A111

\bibitem[{{Str{\"u}der} {et~al}\mbox{.}(2001){Str{\"u}der}, {Briel}, {Dennerl},
  {Hartmann}, {Kendziorra}, {Meidinger}, {Pfeffermann}, {Reppin}, {Aschenbach},
  {Bornemann}, {Br{\"a}uninger}, {Burkert}, {Elender}, {Freyberg}, {Haberl},
  {Hartner}, {Heuschmann}, {Hippmann}, {Kastelic}, {Kemmer}, {Kettenring},
  {Kink}, {Krause}, {M{\"u}ller}, {Oppitz}, {Pietsch}, {Popp}, {Predehl},
  {Read}, {Stephan}, {St{\"o}tter}, {Tr{\"u}mper}, {Holl}, {Kemmer}, {Soltau},
  {St{\"o}tter}, {Weber}, {Weichert}, {von Zanthier}, {Carathanassis}, {Lutz},
  {Richter}, {Solc}, {B{\"o}ttcher}, {Kuster}, {Staubert}, {Abbey}, {Holland},
  {Turner}, {Balasini}, {Bignami}, {La Palombara}, {Villa}, {Buttler},
  {Gianini}, {Lain{\'e}}, {Lumb}, \& {Dhez}}]{struder01}
{Str{\"u}der} L. {et~al.}, 2001, \aap, 365, L18

\bibitem[{{Sugizaki} {et~al}\mbox{.}(2001){Sugizaki}, {Mitsuda}, {Kaneda},
  {Matsuzaki}, {Yamauchi}, \& {Koyama}}]{sugizaki01}
{Sugizaki} M., {Mitsuda} K., {Kaneda} H., {Matsuzaki} K., {Yamauchi} S.,
  {Koyama} K., 2001, \apjs, 134, 77

\bibitem[{{Swartz} {et~al}\mbox{.}(2004){Swartz}, {Ghosh}, {Tennant}, \&
  {Wu}}]{swartz04}
{Swartz} D.~A., {Ghosh} K.~K., {Tennant} A.~F., {Wu} K., 2004, \apjs, 154, 519

\bibitem[{{Swartz} {et~al}\mbox{.}(2011){Swartz}, {Soria}, {Tennant}, \&
  {Yukita}}]{swartz11}
{Swartz} D.~A., {Soria} R., {Tennant} A.~F., {Yukita} M., 2011, \apj, 741, 49

\bibitem[{{Szostek} \& {Zdziarski}(2008)}]{szostek08}
{Szostek} A., {Zdziarski} A.~A., 2008, \mnras, 386, 593

\bibitem[{{Szostek} {et~al}\mbox{.}(2008){Szostek}, {Zdziarski}, \&
  {McCollough}}]{szm08}
{Szostek} A., {Zdziarski} A.~A., {McCollough} M.~L., 2008, \mnras, 388, 1001

\bibitem[{{Tully} {et~al}\mbox{.}(2009){Tully}, {Rizzi}, {Shaya}, {Courtois},
  {Makarov}, \& {Jacobs}}]{tully09}
{Tully} R.~B., {Rizzi} L., {Shaya} E.~J., {Courtois} H.~M., {Makarov} D.~I.,
  {Jacobs} B.~A., 2009, \aj, 138, 323

\bibitem[{{Turner} {et~al}\mbox{.}(2001){Turner}, {Abbey}, {Arnaud},
  {Balasini}, {Barbera}, {Belsole}, {Bennie}, {Bernard}, {Bignami}, {Boer},
  {Briel}, {Butler}, {Cara}, {Chabaud}, {Cole}, {Collura}, {Conte}, {Cros},
  {Denby}, {Dhez}, {Di Coco}, {Dowson}, {Ferrando}, {Ghizzardi}, {Gianotti},
  {Goodall}, {Gretton}, {Griffiths}, {Hainaut}, {Hochedez}, {Holland},
  {Jourdain}, {Kendziorra}, {Lagostina}, {Laine}, {La Palombara}, {Lortholary},
  {Lumb}, {Marty}, {Molendi}, {Pigot}, {Poindron}, {Pounds}, {Reeves},
  {Reppin}, {Rothenflug}, {Salvetat}, {Sauvageot}, {Schmitt}, {Sembay},
  {Short}, {Spragg}, {Stephen}, {Str{\"u}der}, {Tiengo}, {Trifoglio},
  {Tr{\"u}mper}, {Vercellone}, {Vigroux}, {Villa}, {Ward}, {Whitehead}, \&
  {Zonca}}]{turner01}
{Turner} M.~J.~L. {et~al.}, 2001, \aap, 365, L27

\bibitem[{{van den Berg} {et~al}\mbox{.}(2012){van den Berg}, {Penner}, {Hong},
  {Grindlay}, {Zhao}, {Laycock}, \& {Servillat}}]{vandenberg12}
{van den Berg} M., {Penner} K., {Hong} J., {Grindlay} J.~E., {Zhao} P.,
  {Laycock} S., {Servillat} M., 2012, \apj, 748, 31

\bibitem[{{Vilhu} \& {Hannikainen}(2013)}]{vilhu13}
{Vilhu} O., {Hannikainen} D.~C., 2013, \aap, 550, A48

\bibitem[{{Walton} {et~al}\mbox{.}(2013){Walton}, {Fuerst}, {Harrison},
  {Stern}, {Bachetti}, {Barret}, {Bauer}, {Boggs}, {Christensen}, {Craig},
  {Fabian}, {Grefenstette}, {Hailey}, {Madsen}, {Miller}, {Ptak}, {Rana},
  {Webb}, \& {Zhang}}]{walton13}
{Walton} D.~J. {et~al.}, 2013, \apj, 779, 148

\bibitem[{{Warner}(2003)}]{warner03}
{Warner} B., 2003, {Cataclysmic Variable Stars}. Cambridge, UK: Cambridge
  University Press

\bibitem[{{Weisskopf} {et~al}\mbox{.}(2004){Weisskopf}, {Wu}, {Tennant},
  {Swartz}, \& {Ghosh}}]{weisskopf04}
{Weisskopf} M.~C., {Wu} K., {Tennant} A.~F., {Swartz} D.~A., {Ghosh} K.~K.,
  2004, \apj, 605, 360

\bibitem[{{White} \& {Holt}(1982)}]{white82}
{White} N.~E., {Holt} S.~S., 1982, \apj, 257, 318

\bibitem[{{Wilms} {et~al}\mbox{.}(2000){Wilms}, {Allen}, \& {McCray}}]{wilms00}
{Wilms} J., {Allen} A., {McCray} R., 2000, \apj, 542, 914

\bibitem[{{Winter} {et~al}\mbox{.}(2006){Winter}, {Mushotzky}, \&
  {Reynolds}}]{winter06}
{Winter} L.~M., {Mushotzky} R.~F., {Reynolds} C.~S., 2006, \apj, 649, 730

\bibitem[{{Yang} {et~al}\mbox{.}(2009){Yang}, {Wilson}, {Matt}, {Terashima}, \&
  {Greenhill}}]{yang09}
{Yang} Y., {Wilson} A.~S., {Matt} G., {Terashima} Y., {Greenhill} L.~J., 2009,
  \apj, 691, 131

\bibitem[{{Zacharias} {et~al}\mbox{.}(2010){Zacharias}, {Finch}, {Girard},
  {Hambly}, {Wycoff}, {Zacharias}, {Castillo}, {Corbin}, {DiVittorio}, {Dutta},
  {Gaume}, {Gauss}, {Germain}, {Hall}, {Hartkopf}, {Hsu}, {Holdenried},
  {Makarov}, {Martinez}, {Mason}, {Monet}, {Rafferty}, {Rhodes}, {Siemers},
  {Smith}, {Tilleman}, {Urban}, {Wieder}, {Winter}, \& {Young}}]{zacharias10}
{Zacharias} N. {et~al.}, 2010, \aj, 139, 2184

\bibitem[{{Zampieri} \& {Roberts}(2009)}]{zampieri09}
{Zampieri} L., {Roberts} T.~P., 2009, \mnras, 400, 677

\bibitem[{{Zdziarski} {et~al}\mbox{.}(2012){Zdziarski}, {Maitra}, {Frankowski},
  {Skinner}, \& {Misra}}]{zdziarski12}
{Zdziarski} A.~A., {Maitra} C., {Frankowski} A., {Skinner} G.~K., {Misra} R.,
  2012, \mnras, 426, 1031

\bibitem[{{Zdziarski} {et~al}\mbox{.}(2013){Zdziarski}, {Miko{\l}ajewska}, \&
  {Belczy{\'n}ski}}]{zdziarski13}
{Zdziarski} A.~A., {Miko{\l}ajewska} J., {Belczy{\'n}ski} K., 2013, \mnras,
  429, L104

\bibitem[{{Zdziarski} {et~al}\mbox{.}(2010){Zdziarski}, {Misra}, \&
  {Gierli{\'n}ski}}]{zdziarski10}
{Zdziarski} A.~A., {Misra} R., {Gierli{\'n}ski} M., 2010, \mnras, 402, 767

\bibitem[{{Ziosi} {et~al}\mbox{.}(2014){Ziosi}, {Mapelli}, {Branchesi}, \&
  {Tormen}}]{ziosi14}
{Ziosi} B.~M., {Mapelli} M., {Branchesi} M., {Tormen} G., 2014, \mnras, 441,
  3703

\end{thebibliography}

\bsp
\label{lastpage}

\end{document}